\documentclass[print]{mn2e}

\newcommand{\feh}{\mbox{[Fe/H]}}

\usepackage{graphicx}
\title[Photometric properties of Local Volume dwarf galaxies]
{Photometric properties of Local Volume dwarf galaxies\thanks{Based on observations
with the NASA/ESA Hubble Space Telescope, obtained at the Space Telescope Science Institute,
which is operated by the Association of Universities for Research in Astronomy (AURA), 
Inc., under NASA contract NAS 5-26555.}}
\author[M. E. Sharina et al.]
{M. E. Sharina$^{1}$\thanks{E-mail: sme@sao.ru (MES)},
 I. D. Karachentsev$^{1}$, A. E. Dolphin$^{2}$, V. E. Karachentseva$^{3}$, \newauthor R. Brent Tully$^{4}$,
G. M. Karataeva$^{5}$, D. I. Makarov$^{1}$, L. N. Makarova$^{1}$, \newauthor
S. Sakai$^{6}$, E. J. Shaya$^{7}$, E. Yu. Nikolaev$^{8}$, A. N. Kuznetsov$^{5}$\\
$^{1}$Special Astrophysical Observatory, Russian Academy of Sciences, N.Arkhyz, KChR, 369167, Russia\\
$^{2}$Steward Observatory, 933 N. Cherry Ave., Tucson, AZ 85721, USA\\
$^{3}$Astronomical Observatory of Kiev University, Kiev 04053, Ukraine\\
$^{4}$Institute for Astronomy, University of Hawaii, Honolulu, HI 96822, USA\\
$^{5}$Astronomical Institute, St. Petersburg State University, Universitetskii pr. 28, St. Petersburg, 198504, Russia\\
$^{6}$Division of Astronomy and Astrophysics, University of California, at Los Angeles, Los Angeles, CA 90095-1562, USA\\
$^{7}$Astronomy Department, University of Maryland, College Park, MD 20743, USA\\
$^{8}$Department of Astronomy, Kazan State University, Kremlevskaya Street 18, 420008 Kazan, Russia}
\begin{document}

\date{Accepted ---. Received ---}

\pagerange{\pageref{firstpage}--\pageref{lastpage}} \pubyear{2007}

\maketitle

\label{firstpage}
\begin{abstract}
We present surface photometry and metallicity measurements
for 104 nearby dwarf galaxies imaged  with the Advanced Camera for Surveys
and Wide Field and Planetary Camera 2 aboard the Hubble Space Telescope.
 In addition, we carried out photometry for 26 galaxies of the sample and for Sextans~B on images of the Sloan Digital Sky Survey.
Our sample comprises dwarf spheroidal, irregular and transition type galaxies
located within $\sim$10 Mpc in the field and in nearby groups: M81, Centaurus~A,
Sculptor, and Canes Venatici I cloud.
It is found that the early-type galaxies have on average higher metallicity
at a given luminosity in comparison to the late-type objects.
Dwarf galaxies with $ M_B \geq -12 \div -13^m$ deviate toward larger scale lengths
from the scale length -- luminosity
relation common for spiral galaxies, $h \propto L^{0.5}_B$.
The following correlations between fundamental parameters of the
galaxies are consistent with expectations if there is pronounced gas-loss
through galactic winds:
 1) between the luminosity of early-type dwarf galaxies and the mean metallicity of
constituent red giant branch stars, $ Z \sim L^{0.4}$,
2) between mean surface brightness within the 25$^m\!/\sq \arcsec$ isophote and
the corresponding absolute magnitude in the V and I bands, $ SB_{25} \sim 0.3 M_{25}$,
and 3) between the central surface brightness (or effective surface brightness) and
integrated absolute magnitude of galaxies in the V and I bands,
$ SB_{0} \sim 0.5 M_{L}$, $ SB_{e} \sim 0.5 M_{e}$.
The knowledge of basic photometric parameters for a large sample of dwarf galaxies is
essential for a better understanding of their evolution.
\end{abstract}


\section{Introduction}
Environmental density has been argued to be one of the main factors influencing the evolution of galaxies.
Our neighbourhood is predominantly populated by poor groups, made up of
1 -- 2 massive galaxies and numerous dwarf satellites.
A well-known morphology-density relation between the local density in groups and clusters
and galaxy type (see e.g. Hubble \& Humason 1931, Einasto et al. 1974, Dressler 1980,
Binggeli 1994) is also valid for Local Volume low-mass galaxies.
The present day dwarf galaxies are supposed to be relics
of building blocks that form larger galaxies during the hierarchical merging history of the Universe (White \& Rees 1978).
Such objects with typical masses $< 10^8$M$\odot$ lose gas easily due to their shallow potential wells.
Thus, they are strongly influenced by the effects of morphological segregation.
\begin{figure}
\caption{Distribution of our sample and the Local Group galaxies according to their projected distances from
the nearest massive galaxy. The data were taken from Karachentsev et al. 2004. Roughly 50\% dSphs are located
within the projected radius of ~300 kpc from the nearest massive galaxy.}
\label{morf_distr}
\end{figure}
Dwarf galaxies are usually divided into two main morphological classes depending on
their structure, age of stellar populations and gas content:
early-type (dwarf spheroidal, dSph, and elliptical, dE), and late-type (irregular, dIrr,
and blue compact galaxies, BCD) (Kormendy 1985, Karachentseva et al. 1985, Grebel 1999).
The majority of dSphs and dIrrs are low surface brightness (LSB) galaxies with
$(SB)_{0,B} \ge 22.5 \div 23^m\!/\sq \arcsec$, according to the definitions outlined by
Impey and co-authors (Impey \& Bothun 1997, Impey et al. 2001).
The prototypes of dSphs are Sculptor and Fornax dSphs in the Local group (LG).
  There is no sharp difference between dEs and dSphs.
However, dSphs are supposed to have on average lower surface brightnesses and surface brightness 
gradients than dEs (Karachentseva et al. 1985).
The Large and Small Magellanic Clouds can serve as representatives of the brightest dIrrs.
BCDs are systems dominated by very active star formation.
A list of the LG dwarf galaxies and a detailed description of their properties can be found in
a monograph by van den Bergh (2000) and a review by Mateo (1998).

The distinction between different morphological classes is slightly blurred.
New observational data obtained at high spatial resolution and in different wavelength ranges
reveal some peculiar properties of dwarf galaxies.
For example, KDG61 in the M81 group classified as dSph (Karachentseva et al. 1985) contains
a bright HII region and some bright blue stars (Johnson et al. 1997).
Population gradients exist in the Local Group
dSphs (Harbeck et al. 2001) and
in the dwarf elliptical satellites of M31: NGC205, NGC185, É NGC147
(see e.g. Davidge 2005, Sharina et al. 2006 and references therein).
Since dIrrs and dSphs both have exponential surface brightness (SB) profiles,
they are considered to be the members of a common evolutionary sequence.
However, it should be noted that reliable and consistent SB measurements are
complicated for faint galaxies.
So-called transition type galaxies are probably
faded dIrrs, depleted of gas. 

In this paper we concentrate on the properties of dSphs and dIrrs.
The LV dSphs typically are found within the distance of $ \sim$300 kpc around massive
galaxies, while dIrrs are distributed  more uniformly (see Fig.~1).
They occur in groups of galaxies and in the field 
(see also Grebel et al. 2003, and Karachentsev et al. 2005).

Many physical mechanisms are employed to explain the evolution of low-mass galaxies:
ram pressure stripping and stripping by the intergalactic medium,
tidal interactions in groups and clusters, gas blowout through supernova driven wind, etc.
(see e.g. Grebel et al. 2003, Boselli \& Gavazzi 2006, Gunn \& Gott 1972,
Zasov \& Karachentseva 1979, Gnedin 2003, Dekel \& Silk 1987, Ferrara \& Tolstoy, 2000).
Low-mass disks are particularly sensitive to harassment by galaxy encounters and
tidal interactions in high density environments.

Good data that constrain fundamental parameters for
a representative sample of nearby dwarf galaxies would help toward
understanding how galaxies are formed and evolve.
The Hubble Space Telescope (HST) high resolution images represent the best material
for stellar photometry, and detailed studies of the internal structure of galaxies.
Accurate distances to $\sim 200$ objects in the LV were determined
in four snapshot programs (Karachentsev et al. 2001; 2002a, b, c; 2003a, b, c, d; 2006, 2007; 
hereafter: K01-07).
These data provide us the means to study physical properties of a large sample of galaxies.
A main aim of our study is to elucidate basic scaling relations between fundamental parameters
of nearby LSB dwarf galaxies.
The paper is organised as follows. In Section~2 we describe our
approach to data reduction leading to the derivation of the metallicities
and SB profiles. In Section~3 we discuss
the resulting relations based on the derived fundamental
parameters of dwarf galaxies: the metallicity -- luminosity,
the scale length -- luminosity, and the SB --
luminosity relations. In section 4 we formulate our conclusions.

\section{Data reduction and analysis}
Our sample (see Table~1) consists of 104 dwarf galaxies observed
 with the Wide Field and Planetary Camera 2 (WFPC2)
 using the WF3-fix aperture (proposals GO 8192 and 8601),
and with the Advanced Camera for Surveys (ACS) (proposals 9771 and 10235).
All HST observations were carried out with the filters F606W and F814W, the response characteristics of which
can be transformed to the filters V and I in the Johnson-Cousins system (see Holtzman et al. 1995,
Sirianni et al. 2005 for more details).
The exposure times were 600 seconds in each filter for the WFPC2 sample.
The ACS images were taken with the exposure times of 900 and 1200 seconds
 in the F606W and F814W filters, respectively.

We select small galaxies, which are located almost entirely
 within the corresponding WF3 WFPC2 frames.
Since the snapshot sample is restricted in the distance to galaxies,
this selection criterion naturally leads to a limitation in the
luminosity of the objects of study. About 70\% our sample galaxies are fainter
 than $M_V \sim -15^m$ and have central SB in the B-band $(SB)_0 >22.0^m\!/\sq \arcsec$.

 Stellar photometry and artificial star tests were performed using the HST$\!$~PHOT and
DOLPHOT packages (Dolphin 2000a). A detailed description of all reduction steps can
be found in the corresponding manuals. Correction for charge transfer effects was made
using the prescriptions described in Dolphin (2000b). 
Holtzman et al. (2006) tested different techniques of the point-spread function photometry
and concluded that the photometry produced by
the HST$\!$~PHOT package is the most accurate to date. The authors
describe the advantages of using the program,
one the most important of which is a better quality of aperture corrections.
\subsection{Determination of metallicity}
To determine metallicity of red giant branch (RGB) stars in our sample dwarf galaxies we use stellar
photometry results presented in a series of papers devoted to the investigation of the structure of the LV
by Karachentsev and co-authors published between 2002 and 2007 (K01-07) and by Tully et al. (2006).

In these papers the distances were obtained with a typical accuracy better than 12\% using the method of
Da Costa \& Armandroff (1990) and Lee et al. (1993) from the tip of first-ascent
red giant branch stars. Metallicity of old stars was estimated for
some galaxies from a mean color of the RGB measured at an absolute
magnitude $M_I=-3\fm5$ (Lee et al., 1993): $$\feh = 12.64+12.6 \cdot (V-I)_{-3.5} - 3.3 \cdot(V-I)^2_{-3.5}.$$

In this paper we extend the determination of metallicity in the same way to the larger
sample of objects.  The measured metallicities with the corresponding random and
systematic errors are presented in Table~1.
It should be noted, that the exposure times, spatial resolution, and the resulting
photometric depth are different for ACS and WFPC2 images. So, the approaches to the
data reduction were different for these datasets.
\subsubsection{WFPC2}
According to our selection criterion,  only small galaxies, completely covered by the WFPC2 WF3 chip
field of view were included in the analysis.
This circumstance  naturally leads to the limitation of surface brightness, since
the sample of galaxies imaged in the snapshot programs with the WFPC2 is restricted in distance.
Artificial star experiments show that a 50\% stellar detection rate limit for
LSB dwarf galaxies imaged with the WFPC2 at exposure times of 600 seconds in each filter
is usually reached at $V \sim 26^m$ and $I \sim 25^m$.
In Fig.~2 we demonstrate completeness and mean photometric errors for KK~27 and NGC~3741,
which represent the most and least crowded images.

\begin{figure*}
\voffset=1cm
\caption{V completeness and photometric errors for the most and least crowded WFPC2
images (left: KK27, right: NGC3741).
In the bottom panels, the line shows the mean output minus input magnitudes of artificial stars,
and the error bars show the 1$\sigma$ distribution.}
\label{complt}
\end{figure*}
\begin{figure*}
\voffset=-10cm
\caption{Real and artificial color-magnitude diagrams (CMDs) for 
E294-010 (Dist$=1.92$ Mpc) and KK27 (Dist$=3.98$ Mpc)
(see Section 2.1.1 for details).}
\label{cmds}
\end{figure*}
\begin{figure*}
\voffset=-10cm
\caption{Representative CMDs for metal-poor and metal-rich dwarf galaxies
imaged with the WFPC2 (top): E325-011, and KK217;
and the ACS (bottom): KKH6, and KK197 (see Sec.2.1.1 and Sec.2.1.2 for details).
Empirical loci of the red giant branch for Galactic globular clusters M15, M2, NGC1851, and 47 Tuc
(Lee et al. 1993) are overplotted.}
\label{cmds}
\end{figure*}
\begin{figure*}
\caption{Difference between metallicities obtained using the observed and artificial CMDs
versus the corresponding I-band magnitudes of the RGB tip for our sample dwarf galaxies
imaged with the WFPC2.}
\label{Fe_H_Itrgb}
\end{figure*}
\begin{figure*}
\caption{ Dispersion of RGB color measured at $M_I=-3.5$ versus absolute magnitude in the V-band (column 6, Table~1)
for dSphs (dots) and dIrrs (open circles).}
\label{VIspreads}
\end{figure*}
To study how the $\feh $ uncertainties grow with distance,
 we use artificial star experiments
\footnote{Artificial RGBs are presented at anonymous ftp site:~ftp://ftp.sao.ru/pub/sme/DwGal/artRGBs/}
and the following procedure:

1. Selection of artificial stars with magnitudes and colors resembling those
of RGB stars in the studied galaxies;

2. Choice of input magnitudes and colors for recovered stars corresponding
to those of Galactic globular clusters:  M15 (\feh$=2.17$ dex), M2 (\feh$=-1.58$ dex),
NGC1851 (\feh$=-1.29$ dex), and 47 Tuc (\feh$=-0.71$ dex) according to Lee et al. 1993;

3. Determination of the metallicity of the artificial RGBs using the method
of Lee et al. (1993).

  Note that we did not take into account luminosity functions. 
Artificial stars are evenly distributed according to their magnitudes.
 A scatter and a possible systematic difference between the color of an input RGB and that
constructed by recovered stars was used to study \feh\ uncertainties.
In Fig.~3 real and artificial color-magnitude diagrams (CMDs) are shown for one of the most distant and the nearest
galaxies in our sample: KK27 ($I_{TRGB}=24.10$), and E294-010 ($I_{TRGB}=22.38$).
In Fig.~5 we plot the difference between metallicities obtained using the observed
and artificial CMDs versus the corresponding RGB tip magnitude for the galaxies imaged with the WFPC2.
It is seen that the difference is not larger than 0.15 dex for galaxies located
within 4.5 Mpc ($I_{TRGB} \sim 24.2$ mag). So, only the galaxies within this distance range were selected
for the proper determination of metallicity.

Fig.~4 (top) shows representative CMDs for metal-poor and metal-rich dwarf galaxies
imaged with the WFPC2 (see also Karachentsev et al. 2002b).
E325-011 and KK217 are dIrr and dSph satellites of N5128, respectively. Both galaxies are
located at a distance of $\sim350$ kpc from the center of the Centaurus~A group
(Catalog of Neighbouring Galaxies by Karachentsev et al. 2004, hereafter: CNG).
E325-011 is a factor of twenty more luminous than KK217 ($Log L_B^{E325-011}=7.8 L_{\sun}$,
$Log L_B^{KK217}=6.5 L_{\sun}$).
However, it is a factor of $\sim$7 less metal rich ($\feh^{E325-011}=-2.2$ dex, $\feh^{KK217}=-1.2$ dex).
The correspondence between luminosity and metallicity of dwarf galaxies of different
morphological types will be discussed in Sec~3.1.
\begin{figure*}
\caption{Representative surface brightness profiles for dwarf galaxies KK65 and UGC4115,
obtained using three types of images: WFPC2, ACS, and SDSS. Internal photometric errors are overplotted.}
\label{test}
\end{figure*}
\begin{figure*}
\caption{Comparison of properties of dwarf galaxies obtained by us
using the HST images with the corresponding data from the literature (circles)
and our photometry on the SDSS images (see Table) ("s" -- signs).
Small symbols refer to the WFPC2 data, whereas the larger ones represent the ACS data.
Star-like symbols indicate cross-check between our results coming from the WFPC2 and ACS images.
In this case the ACS data are shown along the X axis.
 The panels show the following data obtained using the images in the V and I bands:
integrated magnitudes, and colors (panels 1-3),
effective radii, and SBs (panels 4-7),
model exponential scale lengths and central SBs (panels 8-11),
mean SBs measured within the isophote of 25$^m\!/\sq \arcsec$,
the corresponding V-band radii, and integrated magnitudes (panels 12-16).
All magnitudes and SBs were corrected for Galactic extinction (Schlegel et al. 1998).}
\label{compar}
\end{figure*}
\subsubsection{ACS}
Completeness plots for the dwarf galaxies ESO~269-058 and KKs~55
represent typical most and least crowded fields
imaged with the ACS (see also Karachentsev et al. 2007).
These plots indicate that the 50\% photometric detection limit on the ACS images
is fainter than that of the WFPC2 images.
To avoid possible systematic errors in determination of metallicity we select galaxies
within $ \sim$6 Mpc ($I_{TRGB} \sim 25^m$) for metallicity determination  with the ACS.

In Fig.~4 (bottom) we show representative CMDs for metal-poor and metal-rich dwarf galaxies
imaged with the ACS. High-resolution HST images for these objects were presented by Karachentsev et al. 2006, 2007.
KKH6 is a very isolated metal-poor Irr in the outskirts of the Maffei group ($\feh=-2.4$ dex).
KK197 is an early-type satellite of NGC5128, situated very near to the center of the Centaurus~A group.
The CMD of KK197 shows a wide range of metallicities of RGB stars.
Probably, this is an indication of unusual evolutionary history.
The RGB is very populated, and the mean metallicity at the level $M_I=-3\fm5$ ($\feh=-1.2$ dex) is determined
with small random and systematic errors (Table~1). Both galaxies, KKH6 and KK197,
have similar luminosities $L_B \sim 10^7 L \sun$.
Discussion about the relation between
luminosity and metallicity of dwarf galaxies will be presented in Sec.~3.1.

  The dispersions of RGB colors at $M_I=-3.5^m$ were measured for 30 galaxies observed with the ACS camera.
The objects were chosen to be located roughly in the distance range $2.2 \div 5.6$  Mpc, so that the color
 spreads could be confidently detected. Selection of more distant galaxies leads to a situation where
 the dispersion of RGB colors is comparable with stellar photometry errors. Fluctuations of
RGB color dispersions become prominent for nearer low-mass galaxies. RGBs appear to be insufficiently
populated to measure the color spreads confidently.
The color spread is defined as the square root of the difference between the observed color dispersion
and the corresponding uncertainties of the color estimates taken in quadratures. The color measurement error  
is the sum of V and I measurement uncertainties added in quadrature and scaled according to the transformations 
from  instrumental  F606W, F814W to standard Johnson-Cousins system. Artificial star experiments allow us to 
estimate how crowding influences the uncertainties of stellar magnitudes. Our experiments for the least and the most
crowded images show that the contribution from crowding is insignificant in the case of low surface brightness
dwarf galaxies, i.e. less then $0.03^m$. The resulting dispersions of RGB colors versus absolute $M_{V_0}$
 magnitudes of the galaxies from Table~1
are shown in Fig.~6. The RGB color spread for the giant elliptical galaxy NGC5128 (Rejkuba et al. 2004) is shown versus its
 absolute magnitude from HYPERLEDA (Paturel et al. 2003)~\footnote{http://leda.univ-lyon1.fr}. 
 The dependence is almost linear, and RGB color spreads
are larger for brighter galaxies. The dispersion of colors of RGB stars is influenced by age and metallicity effects.
Detailed inspection of age and metallicity spreads is outside the scope of our paper. However,
one may conclude from inspection of Fig.~6 that star formation events are more frequent and powerful in larger
galaxies, and probably larger galaxies evolve faster. All these facts may be a consequence of the Schmidt star
 formation law (Schmidt 1959)
in the sense that average gas densities are higher in larger galaxies. Our noteworthy case is KK197. 
This nearest dwarf spheroidal satellite of NGC5218 has
twice as high a RGB color dispersion for its luminosity. This is probably a signature of a significant mass loss.
\subsection{Surface photometry}
\subsubsection{Main recipe}
Excellent spatial resolution of HST images allows us to reject background and
foreground objects with high confidence and to obtain detailed data about the structure
of inner parts of dwarf galaxies. The surface photometry was made using
the SURFPHOT program in the MIDAS package developed by ESO.
Major photometry steps were the same for all type of images, and identical
to those used by Makarova (1999).
The background on the cosmic ray cleaned images were determined by
fitting a plane with the FIT/BACKGROUND program.
One of the most important tasks is finding galactic centers.
The centers were usually determined by the ellipse fitting routine FIT/ELL3.
This procedure is complicated for some galaxies, because their
irregular structure consists of multiple bright star-forming regions (for example, DDO53).
In these cases the centers of symmetry were used for the subsequent integration
of light within circular apertures.
Bright foreground stars and background objects were removed from the frames
before integrating the light. Elongated galaxies were processed using elliptical
isophotes.
The majority of our sample galaxies have regular elliptical isophotes and exponential SB profiles
extending to $\sim 27 \div 28^m\!/\sq \arcsec$ in the V-band.
Some galaxies have complex structure with a central depression of light.
Sometimes, bright star-forming regions twist SB profiles.

 We fit the surface brightness profiles of all our sample dwarf galaxies by
an exponential intensity law (de Vaucouleurs 1959) which implies a surface brightness distribution
$$ \mu(r) = \mu_0 +1.086 \cdot (r/h),$$
where $ \mu_0$ is the central surface brightness and h is the exponential scale length.

Our photometric results are presented at an anonymous ftp site~\footnote{
ftp://ftp.sao.ru/pub/sme/DwGal/ACSsbProfiles, ftp://ftp.sao.ru/pub/sme/DwGal/WFPC2sbProfiles,
ftp://ftp.sao.ru/pub/sme/DwGal/SDSSsbProfiles}.
Fundamental photometric properties of our sample objects are listed in Table~1.

Table~1 contains the following data:
{\bf (1)} Galaxy Name and morphological type (second line) from CNG in numerical code
according to de Vaucouleurs et al. (1991),
{\bf (2)} equatorial coordinates (J2000) from CNG,
{\bf (3)} distance modulus (first line), and Galactic extinction in V-band from Schlegel et al. (1998),
{\bf (4)} mean metallicity of the RGBs, {\bf and dispersion of RGB color, $\sigma (V-I)$, measured at $M_I=-3.5$ (first line)} 
with random and systematic errors of mean metallicity of the RGBs separated by a comma (second line),
{\bf (5)} projected major axis from CNG and limiting diameter of our photometry,
{\bf (6)} apparent V and I magnitudes integrated within the diameter $D_{lim}$,
{\bf (7)} limiting V and I surface brightnesses (SB) taken at the distance D$_{lim}$/2 from centers of galaxies,
{\bf (8)} effective $(V\!-\!I)$ color and effective SB in the V-band (superscript "a" indicates an average color),
{\bf (9)} mean SB within the isophote of 25$^m\!/\sq \arcsec$ in the V and I bands,
{\bf (10)} apparent V and I magnitudes integrated within the isophote level 25 $^m\!/\sq \arcsec$.
{\bf (11)} projected major axis radius at the isophote level 25$^m\!/\sq \arcsec$ in the V and I bands
{\bf (12)} effective radius, and model exponential scale length,
{\bf (13)} best exponential fitting central SB in V and I bands and corresponding errors,
Superscripts near the galaxy names indicate the corresponding name of the HST
proposal: "1" refers to the proposals ID 9771 and 10235,
"2" refers to the proposal ID 10210, and "3" refers to the proposals ID 8192 and
8601. Superscript "s" indicates, that SDSS images were used. The data listed in the columns (7)-(11), 
(14) were not corrected for Galactic extinction.

Representative SB profiles are shown in Fig.~7.
In the following we specify the WFPC2 and ACS reduction steps.
\subsubsection{WFPC2}
Surface photometry of galaxies on the WFPC2 HST images was carried out using the recipes
and transformations into the standard Johnson-Cousins system by Holtzman et al. (1995).
All frames  were corrected for distortion.

For our purpose we select galaxies with the CNG angular sizes less than the
corresponding CCD images.
In general, selection of small objects allows us to be sure that structural
and photometric parameters are determined correctly.
We list integrated magnitudes in Table~1 (column 6) only for galaxies located completely in their frames.
\subsubsection{ACS}
\begin{figure*}
\caption{ Luminosities of dwarf galaxies versus mean metallicities of their RGB stars.
\feh\ was estimated from the mean color of the RGB measured at an absolute magnitude$M_I=-3.5$
(Lee et al. 1993). The data for the LG dwarf galaxies were taken from Grebel et al. (2003). 
The data shown for our sample galaxies are listed in Table~1 (Col.~4,~6), while luminosities in the B-band were taken from CNG.}
\label{Fe_H_Lb}
\end{figure*}
\setcounter{figure}{8}
\begin{figure*}
\caption{continued.}
\label{Fe_H_Lb}
\end{figure*}
\begin{figure*}
\caption{
Surface brightness -- luminosity relation in the V and I bands for our sample of dwarf galaxies:
{\bf top:} between mean SBs within the isophote of 25$^m\!/\sq \arcsec$
and the corresponding absolute magnitudes;
{\bf medium}: between central SBs and the absolute magnitudes integrated within the diameters $D_{lim}$ 
(column~6 of Table~1); 
{\bf bottom}: between effective SBs and the corresponding absolute magnitudes.
All data were corrected for Galactic extinction using Schlegel et al. (1998) maps.
 The top and bottom right plots translate to stellar mass from 
the I band. The M/L ratios were estimated
using $(V-I)_0$ colors of the galaxies integrated within the effective radius and
the isophote $25^{m}\!/\sq \arcsec$ (Table~1), Bruzual \& Charlot (2003) SSP models, 
and a Salpeter IMF (Bell \& de Jong, 2001). The symbols are the same as in Fig.~9.}
\label{SB}
\end{figure*}

\begin{figure*}
\caption{Logarithm of the scale length
versus the absolute B (from CNG) and I magnitudes integrated within
the diameters $D_{lim}$ (column~6 of Table~1). The symbols for our sample galaxies
are the same as in Fig.~9. The data for disky galaxies from Vennik et al. (1996),
Binggeli \& Cameron (1993), and van der Kruit (1987) are shown  by squares, plus-signs,
and crosses (x), correspondingly. I magnitudes for disky galaxies were taken from Springob et al. (2007).}
\label{SB0_L}
\end{figure*}
We employ the ACS images processed with the Dither
package\footnote{http://www.stsci.edu/hst/acs/proposing/dither}.
To transform our ACS surface photometry results into the standard Johnson-Cousins
system we use the zeropoints and calibration coefficients from Sirianni et al. (2005).
The 50 pixel inter-chip gap represents a difficulty for the ACS data analysis,
because it often projects onto the central parts of galaxies.
To solve the problem, we interpolate the image intensities within the gap.
Towards this aim we extract parallel strips from the image area
contiguous with the strip and insert them into the gap area.
To test the influence of the inter-chip gap on the derived SB profiles,
we compare our results with the literature data, and with our WFPC2 photometry
results available for some galaxies.
 Additionally, we carried out surface photometry of our sample objects
on the Sloan Digital Survey (SDSS). We found SDSS images for
22 galaxies from the ACS and WFPC2 samples.
The comparison of our ACS surface photometry results with those presented in the literature
and obtained by us using the WFPC2 and SDSS images shows that the fundamental parameters
obtained by us using different types of images are trustworthy (see Fig.~8).
\subsubsection{SDSS}
Large scale, well-calibrated, deep SDSS images provide an excellent homogeneous
material for high-quality photometry.
Surface photometry was made following the steps described in Section~2.2.1.
To transform our measurements into the standard Johnson-Cousins
system, the empirical color transformations by Jordi et al. (2006) were employed.
Extinction coefficients, airmasses, exposure times, and other important data
were taken for each galaxy from the corresponding field table entries.

In Fig.~8 the comparison of the fundamental photometric parameters derived
using the ACS, WFPC2, and SDSS images is demonstrated.
It is seen, that the typical accuracies of the 
 V and I magnitudes integrated within the limiting diameter $D_{lim}$ (Table~1)
and isophote 25$^m\!/\sq \arcsec$, mean SBs within  the isophote 25$^m\!/\sq \arcsec$, effective SBs,
and $(V\!-I)$ color are typically $\le 0\fm2$.
Exponential scale lengths and effective radii are obtained with typical
uncertainties of about 2$\arcsec$ for small galaxies ($R_e \leq 30\arcsec$).
For larger galaxies uncertainties of this model parameters can reach $\sim$30\%.
There are no systematical trends in estimation of the aforementioned
photometric parameters. We list apparent magnitudes integrated
within the limiting diameter $D_{lim}$ in Table~1 (column 6) only if the limiting diameter of our photometry
is greater or equal than the major axis of a galaxy from CNG. However, it would be incorrect to
interpret these magnitudes as total magnitudes. For many our sample galaxies a ratio between a limiting radius
from a center of a galaxy to a disk scale length is less than two. This means that if we assume exponential
 surface brightness profiles for our sample galaxies, the
difference between the corresponding extrapolated total magnitude and an isophot limited magnitude
may be greater than $0.5^m$ according to an equation (7) from Tully et al. (1996): $\delta m = 2.5~log~[1 - (1+r/h) e^{-r/h}],$
where r is a limiting radius of surface photometry, and h is an exponential scale length.

In Table~2 we present fundamental parameters for our HST-based sample
of dwarf galaxies imaged with the SDSS. Additionally, we include
in this table the data for the nearby dwarf irregular galaxy Sextans~B.
  Successive columns contain the following data:
(1) Galaxy Name. (2) equatorial coordinates (J2000),
(3) projected major axis from CNG and limiting diameter of our photometry,
(4) apparent B and R magnitudes integrated within the limiting diameter,
(5) limiting B and R surface brightnesses (SB) taken at the distance D$_{lim}$/2 from center of galaxy,
(6) effective $(B\!-\!R)$ color and effective SB in the R-band,
(7) mean SB within the isophote of 25$^m\!/\sq \arcsec$ in the B and R bands,
(8) apparent B and R magnitudes integrated within the isophote level 25 $^m\!/\sq \arcsec$,
(9) projected major axis at the isophote level 25$^m\!/\sq \arcsec$ in the B and R bands,
(10) effective radius and model exponential scale length,
(11) central SB in B and R bands and corresponding errors.
The data were not corrected for Galactic extinction.
\subsubsection{The surface photometry errors}
 The magnitude errors are usually divided into internal and external components.
We derived external photometric errors by the comparison of our measurements
with the literature data and by the cross-comparison of photometric
parameters obtained using different types of images.
It was shown in the previous section (see also Fig.~8) that the measured magnitudes
agree with the independently estimated ones typically within $0\fm2$.

Internal components consist of random and systematical errors.
Random uncertainties are mainly defined by the accuracy of the background estimates.
The standard deviation of surface brightness in photons per unit area is $\sqrt{g \cdot I + r^2}$,
where I is the total object and background flux in ADU (analog-to-digital-units) in this area,
g is the gain in electrons per ADU, and r is readout noise in electrons.

The systematic component includes the errors of transformations into the standard system
and uncertainties of internal extinction estimation.
The uncertainty of the WFPC2 photometric zero point is estimated to be within $0\fm05$
(Dolphin 2000b). The errors of empirical coefficients for the transformations from WFC to BVRI
do not exceed 0.08 for the filter F606W and 0.02 for F814W (Sirianni et al., 2005).
Given the similarity of our sample dSphs and dIrrs to the LG galaxies we estimate that
the correction for the internal reddening is $E_{(B-V)}\la 0\fm1$ James et al. (2005).

Summing up the internal errors, we obtain the resulting errors of the integrated magnitude
and the surface brightness estimations to be consistent with the external errors
derived by the comparison of different measurements.
\section{Results}
\subsection{The luminosity -- metallicity relation}
In the top panel of Fig.~9 we show the dependence between B-band luminosities from CNG
and metallicities of RGB stars  for our sample galaxies.  In the next two panels of Fig.~9
the same metallicities are plotted versus luminosities in the V and I bands obtained in our study.
The errors of metallicity estimations are presented in Table~1, the photometry  uncertainties
are discussed in Section~2.2.5.

It has been shown by many authors that the more luminous dwarf galaxies are on average
the most metal rich (Mateo 1998, and references therein; van den Bergh 1999; Grebel et al. 2003).
The luminosity -- metallicity (Z -- L) dependency for the Local Group dwarf galaxies have been
quantified by Dekel \& Silk (1986). Caldwell et al. (1992) confirmed this relation using Mg$_2$
spectral line index for metallicity estimation, and showed that
the Z -- L dependency obtained using this method is steeper ($Z \sim L^{0.6}$)  for more luminous E galaxies.
Thomas et al. (2003) found a unified relation $[Z/H]=-3.6-0.19 M_B$ for
early type dwarf galaxies and for giant ellipticals.
 Grebel et al. (2003) stressed attention to the fact that the LG dSphs have a higher
mean metallicity of old stellar populations for a fixed optical luminosity in comparison with dIrrs.
Fig.~9 shows that a similar tendency holds for our sample of galaxies
situated in the nearby groups and in the field.

There may be several reasons for the separation between dIrr and dSph in the Fig.~9
metallicity--luminosity plots (see also Grebel et al. 2003 and references therein).
First, perhaps the displacement being interpreted as a metallicity effect is
actually an age effect.
The average metallicity derived by
the method of Lee et al., 1993 is only valid for stellar populations with the ages
comparable to that of Galactic globular clusters. The age-metallicity degeneracy
problem has been discussed for a long time
(Worthey 1994, Saviane et al. 2000, Dolphin et al. 2003, Salaris \& Girardi 2005,
see also Zasov \& Sil'chenko 1983).
Fig.~5 from the paper by Dolphin et al. (2003) provides an exhaustive
 explanation of this effect.
Probably, average ages of old stars in dIrrs differ from galaxy to galaxy and
are systematically less than those in dSphs.
The Z -- L relation for normal and dwarf elliptical galaxies
can also be skewed by the presence of a significant intermediate-age component.
However, this effect in dSph is not so pronounced as in the case of dwarf irregular galaxies.

A second explanation for what is being seen in Fig.~9 may lie in the influence of 
environments on the evolution
of dwarf galaxies. The metallicity of interstellar gas is dependent on three factors:
the rate of star formation, the degree of gas expulsion, and the influx of new gas.
It was shown by Shaya \& Tully (1984) that intergalactic material will not fall 
into a galaxy once it is in a group.
Galaxies in isolation can continue to accrete fresh shells of low-metallicity intergalactic gas.
However, the cluster potential creates a Roche limit around an individual galaxy.
Gas envelopes lost by galaxies from violent star forming events are tidally stripped
and acquired by the group or cluster.
Shells of pristine interstellar gas that can fall into an isolated
dIrr and dilute its metal content are
prevented from falling into a dSph in the larger potential well of a
group or cluster.

A third possibility is that the displacement between dIrr and dSph in Fig.~9  is
partially in luminosity. Passive evolution of dIrr galaxies leads to a fading of $ \sim 2^m$ 
of their integrated magnitude if they stop forming stars (Hunter \& Gallagher 1985).
Tidal mass loss by dwarf galaxies located near the centers of groups and clusters should not be ignored.
  Low density galaxies entering a cluster may get completely disrupted 
by tidal shocks over a time scale less than the Hubble time (e.g. Merritt 1984,
Moore at al. 1999. Mayer et al. 2001, Gnedin 2003, Koposov et al. 2007, 
McConnachie et al. 2007).
 However, it is difficult to accept that disruption and fading alone explains the observed offset.
The displacement of one to two orders of magnitude in luminosity implies
a loss of 90\% to 99\% of light for essentially all early-type dwarfs.
An ingredient might be added 
if similar physical mechanisms worked at the stage of dSphs formation. 
The difference in metallicity at the
same luminosity between dSph and dIrr may, at least partially be due to the earlier, faster transformation
of gas into stars in a denser environment (Hogg et al. 2003, Thomas et al. 2005)
and a subsequent fading after the exhaustion of gas and loss of stars.

  Five dwarf galaxies classified as dSphs have low metallicity values at a given luminosity that resemble
those of dIrrs (Fig.~9): NGC5237, KKs55, CenN, KK189, and KK27. They either
have a significant intermediate-age component, or have a gas recycling history similar to that of dIrrs,
perhaps because they are in transition from the dIrr to the dSph state.

Several contradictory observational and theoretical studies regarding the existence of a correlation
between luminosity and metallicity of young stellar populations in dIrrs have been presented in the literature.
Note, that [O/H] abundance is often used as the indication of the metallicity of dIrrs.
For instance, McGaugh 1994, Richer \& McCall (1995), Hidalgo-Gamez et al. (2003) found a weak relationship with luminosity.
Other authors (e.g. van Zee et al. 1997, Skillman et al. 1989) established a strong correlation.
Saviane et al. (2005) interpreted the existence of near-infrared Z -- L relation for dIrrs following
the considerations by Skillman et al. (2003) and Pilugin \& Ferrini (2000).
In the evolutionary closed-box scenario the gas fraction decreases monotonically
as metal abundance increases. Larger galaxies evolve faster.
On the other hand, Hidalgo-Gamez et al. (2003) showed that the relationship between the metal content
of a galaxy and its mass fraction of gas is not always linear.
Those authors concluded that variations in the  stellar mass-to-light ratio can contribute
significantly to the scatter in the Z -- L relation.

According to the model of Dekel \& Silk (1986),
the formation of small, diffuse, metal-poor galaxies is  mainly regulated by supernova-driven winds.
Galaxies with the virial velocity and mass below a critical value undergo substantial
gas loss as a result of a violent burst of star formation.
Luminosity is proportional to metallicity as $L \sim Z^{2.5}$ in this scenario.
One can see that our dependency for early-type dwarf galaxies lies close to the theoretical one.
This result suggests that the efficiency of star formation is regulated by similar mechanisms in
early-type dwarf galaxies of different luminosities. It increases with increasing mass
and potential well of the object. Smaller galaxies lose larger gas fractions through
galactic winds.
\subsection{The surface brightness -- luminosity relation}
The relation between the mean SB of galaxies within the isophote level of
25$^m\!/\sq \arcsec$ corrected for Galactic extinction using maps by Schlegel et al.
(1998), and the corresponding absolute magnitude
integrated within the same isophote is shown in Fig.~10~(top).
The slope of the dependency is well defined: $SB_{V_{25}} \sim 0.33 M_{25}$.
Data drawn from the CNG give the correlation between a mean
SB and an absolute magnitude in the B-band, $SB_B \sim 1/3 M_B$,
for all neighbouring galaxies and imply a constancy of
galactic spatial density averaged within this isophote. In other words, $L \sim R_{25}^3$,
where $R_{25}$ is the radius of the galaxy within the isophote level 25$^m\!/\sq \arcsec$.
Our study shows that this conclusion is valid for SBs and luminosities measured in the V and I bands.
These findings imply that the mean luminosity density within $\sim$10 Mpc is
only weakly dependent on the mass of a galaxy and its environment.

In distinction to the average SB, the central SB
shows a steeper dependence on the integrated luminosity (Table~1, column~6):
 $SB_{0} \sim 0.5 M_{L}$ (see Fig.~10, medium panel).
 This correlation is consistent with the luminosity -- radius relationship, $L \sim R^4$,
 valid for the LG dwarf galaxies (Dekel \& Silk 1986).
It can be explained by a gas outflow model, first proposed by Larson (1974).
Vader (1986) and Dekel \& Silk (1986) applied it to dwarf galaxies.
A small galaxy loses an increased fraction of its mass through stellar winds.
This reinforces the decrease of its gravitational potential and, as a consequence, of the central
surface mass density.

The mean SB inside the circular aperture enclosing one-half
the total flux seems to be sensitive to the mass lost in the same way.
The slope of the "effective SB -- effective absolute magnitude"
relation (Fig.~10, bottom) coincides with the previous one: $SB_{V_{e}} \sim 0.5 M_{e}$.

The scatter in surface brightness at a given luminosity can be not only the result of
photometric errors and approximation uncertainties, but also may be influenced
by physical processes. For example, surface brightness can depend on angular momentum
(Dalcanton et al. 1997).
\subsection{The scale length -- luminosity relation}
Fig.~11 shows the exponential scale length -- luminosity relation (h -- L) for our sample
objects and for some brighter spiral galaxies.  We use the integrated magnitudes from
the column~6 of Table~1 for calculation of the luminosities of dwarf galaxies in the I-band.
It is seen that scale length is proportional to luminosity as $h \sim L^{0.5}$
in the B and I bands for spiral galaxies brighter than $M_B \sim -12 \div -13^m$.
It is known since a pioneering work of Freeman (1970) that the central surface brightness
of spiral galaxies has a characteristic value, which  means that the total luminosity of
the disk is $L_T = 2 \pi I_0 h^2 \propto h^2$, where $I_0$ is the central brightness.
It is still under debate whether this result is common for all disky galaxies (see e.g. O'Neil \& Bothun, 2000).
The slope of the h -- L relation is identical for our sample galaxies with absolute blue magnitudes
falling in the range of $M_B \sim -16 \div -13$ and for the spiral galaxies
from the samples of Vennik et al. (1996), Binggeli \& Cameron (1993), and van der Kruit (1987).
However, fainter objects deviate from this dependency towards larger scale lengths.
The different slope of the h -- L dependency for faint and bright galaxies is
caused by a prominent  central surface brightness -- luminosity dependency
for dwarf galaxies shown above.
Passing from bright to faint, luminosities at first decrease with $I_0 \sim constant$
because h decreases, then continue to decrease because $I_0$ decreases with h
only weakly decreasing.
The change in the slope of the h -- L relation is presumably caused by the fact that
galaxies with masses below some transition mass lose their gas more efficiently.
It was shown theoretically by Mac Low \& Ferrara 1999, Ferrara \& Tolstoy 2000, Marcolini et al. 2006,
and  Read et al. 2006 that this limit corresponds to a critical mass $M_{crit}\sim 10^8 M_{\sun}$
in the case of mass-loss through stellar winds. This limit roughly corresponds
$M_B \sim -12 \div -13^m$ (Bell \& de Jong, 2005).

We have no means to quantify the weak h -- L dependency for faint galaxies.
First, our data are not sufficiently numerous.
Second, sizes and luminosities of dwarf galaxies may vary with the environmental density.
Gnedin (2003) used high-resolution cosmological simulations to explore
tidal heating of high and low-mass disks in clusters. He found that the possibility
that a galaxy entering a cluster will be completely disrupted by tidal shocks depends on
interplay between surface brightness of the object and its scale length
in the sense that more compact and luminous galaxies survive longer. Disks of large spiral galaxies
are expected to be thicken by a factor of 2 or 3 in dense environments. Disks of LSB dwarf galaxies
are even more affected by tidal heating.
\section{Conclusions}
In this paper we study the basic photometric properties of 104 dwarf galaxies situated
within 10 Mpc in the field and in nearby groups. Surface photometry and the
determination of mean metallicity of old stellar populations was made for
the majority of the galaxies for the first time. Scaling relations found by us are in line
with the model of dwarf galaxy formation as a result of supernova-driven winds.
We realize, that this interpretation is rather simplistic, and should be viewed with caution.

Using our stellar and surface photometry results we obtain relations between metallicity,
structural, and photometric properties of LSB dwarf galaxies.
We suggest that effects theoretically predicted in the literature and observationally tested and
quantified in our paper: the "metallicity -- luminosity", the "scale length -- luminosity", and the
"surface brightness -- luminosity" relations, may be the consequences of
gas loss by dwarf galaxies through supernova-driven winds, which is dependent on the mass of galaxies
and their environments.

The data on the luminosity and metallicity for our sample objects correspond well 
with the known properties of the Local Group dwarf galaxies.
Apparently, the processes that govern the evolution of dwarf galaxies are similar within the LV,
populated mainly by poor groups. The $Z \sim L^{0.4}$ relation for our sample objects
is consistent with those predicted theoretically by Dekel \& Silk (1986) for dwarf galaxies, and with
the literature information on the faint end of the general relations for early-type galaxies of different masses.

The dSphs have higher apparent metallicity RGB stars as determined using the method of Lee et al. (1993)
than dIrrs at a given luminosity. This fact is probably caused by younger ages of 
evolved stellar populations in
dIrrs in comparison with dSphs, and by environmental differences in accretion rates 
of pristine interstellar gas between dIrr and dSph
dwarf galaxies.

The relation between the surface brightness averaged within the isophote of 25$^m\!/\sq \arcsec$ and luminosity
for our sample objects, $SB_{25} \sim0.33 M_{25}$, implies the constancy of this averaged spatial luminosity density
from giant to dwarf galaxies.

The central and effective SB -- luminosity relations in the V and I bands, $SB_{0} \sim 0.5 M_{L}$ and
$SB_{e} \sim 0.5 M_{e}$, are consistent with the luminosity -- radius relation, $L \sim R^4$.
It was shown by Dekel \& Silk (1986) that a significant gas loss in a dominant dark matter halo with
upper halo circular velocity of $\sim 100$ km s$^{-1}$
had to take place to reproduce this observational result.

We extend the observational scale length -- luminosity relation to faint magnitudes.
Galaxies fainter than $M_B \sim-13^m \div -12^m$ systematically deviate towards larger scale lengths
from the dependency $h \propto L^{0.5}$, valid for bright spiral galaxies.
We interpret this fact as a signature of the environmentally dependent mass-loss in small galaxies.
Galaxies less massive than $ 10^8 M_{\sun}$ lose their gas efficiently through galactic winds.
Tidal interactions between galaxies and the interaction with the intergalactic medium
additionally influence the structure of early-type dwarfs located near the centers of groups and clusters.
\section*{Acknowledgments}
Support associated with HST programs 8192, 8601, 9771, 10219, and 10235
was provided by NASA through a grant from the Space Telescope Science
Institute. Data was extracted from the
the SDSS Archive has been provided by the Alfred P. Sloan Foundation,
the Participating Institutions, The NASA, NSF, DOE,
Japanese Monbukagakusho, and the Max Planck Society. We thank Mario Mateo for helpful comments.
MES, IDK, VEK, DIM, and LNM greatfully acknowledge a partial financial support of the grants: RFBR 07-02-00005
and DFG-RFBG06-02-04017. GMK acknowledges the RFBR grant 05-02-17548.

\newpage
\begin{table*}
\begin{center}
\scriptsize
\caption{Properties of the Local Volume dwarf galaxies. Columns contain the following data:
{\bf (1)} Galaxy Name and morphological type (second line) from CNG in numerical code
according to de Vaucouleurs et al. (1991),
{\bf (2)} equatorial coordinates (J2000) from CNG,
{\bf (3)} distance modulus (first line), and Galactic extinction in V-band from Schlegel et al. (1998),
{\bf (4)} mean metallicity of the RGBs, $\sigma (V-I)$ of RGBs measured at $M_I=-3.5$ (first line) 
with random and systematic errors of mean metallicity of the RGBs separated by a comma (second line),
{\bf (5)} projected major axis from CNG and limiting diameter of our photometry,
{\bf (6)} apparent V and I magnitudes integrated within the diameter $D_{lim}$,
{\bf (7)} limiting V and I surface brightnesses (SB) taken at the distance D$_{lim}$/2 from centers of galaxies,
{\bf (8)} effective $(V\!-\!I)$ color and effective SB in the V-band (superscript "a" indicates an average color),
{\bf (9)} mean SB within the isophote of 25$^m\!/\sq \arcsec$ in the V and I bands,
{\bf (10)} apparent V and I magnitudes integrated within the isophote level 25 $^m\!/\sq \arcsec$.
{\bf (11)} projected major axis radius at the isophote level 25$^m\!/\sq \arcsec$ in the V and I bands
{\bf (12)} effective radius, degree of Sersic profile,
{\bf (13)} best exponential fitting central SB in V and I bands and corresponding errors,
Superscripts near the galaxy names indicate the corresponding name of the HST
proposal: "1" refers to the proposals ID 9771 and 10235,
"2" refers to the proposal ID 10210, and "3" refers to the proposals ID 8192 and
8601. Superscript "s" indicates, that SDSS images were used. The data listed in the columns (7)-(11), 
(14) were not corrected for Galactic extinction.
}
\vspace{-0.5cm}
\begin{tabular}{lrrlccclccclc}\\ \hline \hline
Name              &  RA(2000)& $\mu_0$& \feh,$\sigma_{VI}$ &   Diam   & V$_L$& $SBV_{L}$& $(V\!-\!I)_e$& $SBV_{25}$ & $V_{25}$ & $R_{V,25}$& $R_e$& $SBV_{C}$  \\
T          & DEC(2000) & $A_v$   & $ \sigma$\feh  & $D_{lim}$& $I_L$& $SBI_{L}$ & $SBV_{e}$  & $SBI_{25}$ & $I_{25}$ & $R_{I,25}$& h   & $SBI_{C}$  \\
	   &           & $^{m}$& dex,$^{m}$& $\arcmin$& $^{m}$& $^{m}\!/\sq \arcsec$& $^{m},^{m}\!/\sq \arcsec$& $^{m}\!/\sq \arcsec$& $^{m}$& $\arcsec$& $\arcsec$ & $^{m}\!/\sq \arcsec$\\
(1)        &    (2)& (3) & (4) & (5) & (6) & (7) & (8) & (9) & (10)& (11)& (12)& (13)\\
\hline
E349-031$^1$     &   00 08 13.3& 27.48&  -1.76,0.07 &   1.10 & 15.42& 25.7 &  0.67   & 23.81 &  15.77 &  25.56 & 26&  23.03$\pm$0.06 \\
  10             &  -34 34 42.0& 0.04 &  0.03,0.14  &   1.28 & 14.71& 25.0 & 23.81   & 23.35 &  14.80 &  34.13 & 16    &  22.41$\pm$0.06 \\
E410-005$^3$     &   00 15 31.4& 26.42&  -1.93      &   1.30 & 14.88& 24.5 &  1.06   &       &        &        & 30&  22.22$\pm$0.01 \\
  -1             &  -32 10 48.0& 0.058&  0.05,0.15  &   1.12 & 13.93& 23.3 & 23.12   &       &        &        & 18    &  21.43$\pm$0.01 \\
Sc22$^3$         &   00 23 51.7& 28.12&             &   0.90 & 17.47& 26.3 & 1.2     & 24.97 &  19.06 &   9.06 & 35&  24.72$\pm$0.02 \\
  -3             &  -24 42 18.0& 0.05 &             &   0.83 & 16.31& 24.9 &         & 24.26 &  16.31 &  24.90 & 26    &  23.83$\pm$0.02 \\
E294-010$^3$     &   00 26 33.3& 26.42&  -1.48      &   1.10 & 15.29& 24.4 &  0.91   &       &        &        & 25&  22.28$\pm$0.02 \\
  -3             &  -41 51 20.0& 0.02 &  0.04,0.13  &   0.98 & 14.43& 23.5 & 23.34   &       &        &        & 15    &  21.50$\pm$0.03 \\
DDO226$^3$       &   00 43 03.8& 28.46&             &   2.20 &      & 23.7 & 0.84$^a$&       &        &        & 38&  22.60$\pm$0.04 \\
  10             &  -22 15 01.0& 0.05 &             &   1.01 &      & 22.8 &         &       &        &        & 27    &  21.93$\pm$0.03 \\
KDG2$^3$         &   00 49 21.1& 27.66&  -1.75      &   1.20 & 16.12& 26.0 &  1.28   & 24.24 &  16.35 &  26.19 & 30&  23.37$\pm$0.01 \\
  -1             &  -18 04 28.0& 0.08 &  0.06,0.15  &   1.14 & 15.00& 24.6 & 24.27   & 23.42 &  15.00 &  34.06 & 18    &  22.46$\pm$0.01 \\
DDO6$^3$         &   00 49 49.3& 27.62&  -2.11      &   1.70 &      & 24.7 & 0.83$^a$&       &        &        & 36&  23.13$\pm$0.04 \\
  10             &  -21 00 58.0& 0.06 &  0.07,0.16  &   0.99 &      & 23.7 &         &       &        &        & 23    &  22.45$\pm$0.04 \\
E540-032$^3$     &   00 50 24.3& 27.67&  -1.45      &   1.30 & 16.10& 25.6 & 16.13   & 24.30 &  16.32 &  27.59 & 34&  23.54$\pm$0.01 \\
  -3             &  -19 54 24.0& 0.07 &  0.06,0.13  &   1.14 & 14.99& 24.3 & 24.46   & 23.45 &  14.99 &  34.26 & 24    &  22.76$\pm$0.01 \\
UGC685$^2$       &   01 07 22.3& 28.40&   -1.54,0.17&   1.40 & 13.70& 25.3 &  0.86   & 23.22 &  13.74 &  48.15 & 24&  21.37$\pm$0.01 \\
   9             &   16 41 02.0& 0.02 &  0.01,0.13  &   1.75 & 12.86& 24.4 & 22.25   & 22.52 &  12.86 &  52.45 & 15    &  20.66$\pm$0.01 \\
KKH5$^3$         &   01 07 32.5& 28.15&             &   0.60 & 17.05& 25.7 &  1.61   & 24.16 &  17.70 &  14.04 & 20    &  23.54$\pm$0.03 \\
  10             &   51 26 25.0& 0.94 &             &   0.83 & 15.72& 24.3 & 24.53   & 23.49 &  15.72 &  25.00 & 12    &  22.52$\pm$0.02 \\
KKH6$^1$         &   01 34 51.6& 27.90&   -2.38,0.13&   0.80 & 16.27& 25.8 &  1.26   & 24.08 &  16.47 &  19.81 & 20&  23.11$\pm$0.09 \\
  10             &   52 05 30.0& 1.16 &   0.04,0.18 &   0.86 & 15.05& 24.4 & 24.04   & 23.08 &  15.05 &  25.94 & 12    &  22.07$\pm$0.07 \\
KK16$^{2,3}$     &   01 55 20.6& 28.38&   -1.66,0.11&   0.80 & 15.85& 27.3 &  0.93   & 23.95 &  16.31 &  21.20 & 17&  22.79$\pm$0.01 \\
  10             &   27 57 15.0& 0.23 &   0.04,0.14 &   1.55 & 14.90& 26.1 & 23.70   & 23.48 &  15.16 &  29.05 & 11    &  22.07$\pm$0.01 \\
KK17$^{2,3}$     &   02 00 09.9& 28.37&   -1.75,0.10&   0.60 & 16.80& 27.8 &  1.04   & 24.46 &  17.99 &  11.50 & 18&  23.82$\pm$0.01 \\
  10             &   28 49 57.0& 0.18 &   0.07,0.14 &   1.26 & 15.83& 26.4 & 24.77   & 23.96 &  16.28 &  20.40 & 11    &  22.98$\pm$0.01 \\
KKH18$^3$        &   03 03 05.9& 28.23&             &   0.70 & 16.17& 26.2 &  1.12   & 23.69 &  16.35 &  20.62 & 16    &  22.49$\pm$0.02 \\
  10             &   33 41 40.0& 0.66 &             &   0.99 & 15.13& 24.8 &  23.37  & 23.10 &  15.13 &  29.68 & 10    &  21.69$\pm$0.02 \\
NGC1311$^2$      &   03 20 07.4& 28.70&   -1.33     &   3.20 &      & 24.0 &  0.78   &       &        &        & 23&  20.45$\pm$0.01 \\
   9             &  -52 11 06.0& 0.07 &  0.01,0.11  &   1.55 &      & 23.2 & 21.37   &       &        &        & 14    &  19.82$\pm$0.01 \\
KK27$^2$         &   03 21 05.7& 28.00&   -1.72,0.10&   1.20 & 16.44& 27.2 &  0.96   & 24.66 &  17.77 &  13.65 & 25&  24.13$\pm$0.01 \\
  -3             &  -66 19 22.0& 0.25 &  0.03,0.14  &   1.55 & 15.45& 26.6 & 25.01   & 24.11 &  16.02 &  25.10 & 15    &  23.27$\pm$0.01 \\
IC1959$^2$       &   03 33 11.8& 29.03&   -1.80     &   3.00 & 12.77& 27.6 &  0.80   & 22.85 &  12.89 &  58.70 & 26&  20.70$\pm$0.01 \\
   9             &  -50 24 38.0& 0.04 &  0.02,0.15  &   3.22 & 12.02& 26.6 & 21.60   & 22.65 &  12.06 &  74.55 & 16    &  20.00$\pm$0.01 \\
UGCA92$^1$       &   04 32 00.3& 26.28&             &   2.00 & 14.55& 26.8 &  1.78   & 24.71 &  15.54 &  38.95 & 70&  24.18$\pm$0.01 \\
  10             &   63 36 50.0& 2.63 &             &   3.03 & 12.87& 25.0 & 25.08   & 23.73 &  12.87 &  91.00 & 42    &  22.61$\pm$0.01 \\
CamB$^3$         &   04 53 06.9& 27.52&             &   2.20 &      & 26.0 & 1.26$^a$& 24.43 &  17.45 &  25.64 & 29&  23.83$\pm$0.01 \\
   10            &   67 05 57.0& 0.14 &             &   0.89 &      & 24.4 & 24.75   & 23.59 &  15.54 &  26.79 & 17    &  22.82$\pm$0.01 \\
KKH34$^3$        &   05 59 41.2& 28.32&             &   0.90 & 16.37& 25.3 &  0.69   & 24.05 &  16.56 &  21.51 & 22    &  23.22$\pm$0.01 \\
  10             &   73 25 39.0& 0.83 &             &   0.87 & 15.57& 24.7 & 24.13   & 23.46 &  15.57 &  26.19 & 13    &  22.27$\pm$0.02 \\
E121-20$^1$      &   06 15 54.5& 28.86&   -2.39     &   1.40 & 15.36& 27.3 &  0.69   & 23.87 &  15.65 &  26.90 & 19&  22.63$\pm$0.01 \\
  10             &  -57 43 35.0& 0.13 &   0.05,0.18 &   1.55 & 14.65& 26.5 & 23.57   & 23.53 &  14.78 &  34.10 & 12    &  22.08$\pm$0.02 \\
E489-56$^3$      &   06 26 17.0& 28.49&             &   0.60 & 15.49& 25.0 &  0.98   & 23.39 &  15.49 &  26.49 & 17&  21.94$\pm$0.02 \\
  10             &  -26 15 56.0& 0.21 &             &   0.88 & 14.65& 23.9 & 22.86   &       &        &        & 10    &  21.30$\pm$0.01 \\
E490-17$^3$      &   06 37 56.6& 28.13&             &   1.70 &      & 23.4 & 0.90$^a$&       &        &        & 34&  21.92$\pm$0.05 \\
  10             &  -25 59 59.0& 0.26 &             &   0.98 &      & 22.5 &         &       &        &        & 20    &  21.07$\pm$0.03 \\
KKH37$^1$        &   06 47 45.8& 27.62&   -1.80,0.08&   1.20 & 15.39& 28.4 &  0.88   & 23.74 &  15.74 &  24.55 & 17&  22.40$\pm$0.01 \\
  10             &   80 07 26.0& 0.25 &   0.02,0.15 &   1.78 & 14.48& 27.4 & 23.34   & 23.28 &  14.61 &  33.25 & 10    &  21.50$\pm$0.01 \\
UGC3755$^{2,3}$  &   07 13 51.8& 29.35&             &   1.70 & 13.62& 25.0 &  0.98   & 22.49 &  13.62 &  58.97 & 25&  21.27$\pm$0.02 \\
10               &   10 31 19.0& 0.29 &             &   1.97 & 12.75& 24.1 & 21.81   & 21.72 &  12.75 &  58.97 & 15    &  20.58$\pm$0.02 \\
E059-01$^1$      &   07 31 19.3& 28.24&  -1.52,0.18 &   2.10 & 13.69& 25.5 &  0.96   & 23.43 &  14.01 &  49.99 & 48&  22.72$\pm$0.05 \\
   9             &  -68 11 10.0& 0.49 &  0.01,0.13  &   2.80 & 12.70& 24.4 & 23.40   & 22.63 &  12.70 &  83.98 & 29    &  21.72$\pm$0.04 \\
UGC3974$^{2,S}$  &   07 41 55.0& 28.57&             &   3.10 &      & 25.5 &  0.83   & 24.05 &  14.45 &  51.35 & 53&  23.16$\pm$0.01 \\
  10             &   16 48 02.0& 0.11 &             &   2.40 &      & 24.8 & 24.08   & 23.63 &  13.28 &  71.95 & 32    &  22.44$\pm$0.02 \\
KK65$^{2,3,S}$   &   07 42 31.2& 28.27&             &   0.90 & 14.86& 26.1 &  0.77   & 23.50 &  15.09 &  30.20 & 19&  22.00$\pm$0.01 \\
  10             &   16 33 40.0& 0.11 &             &   1.54 & 14.10& 25.3 & 22.86   & 23.29 &  14.16 &  41.30 & 11    &  21.29$\pm$0.01 \\
UGC4115$^{S,3,2}$&   07 57 01.8& 29.44&             &   1.80 & 14.36& 25.1 &  0.88   & 23.53 &  14.41 &  40.79 & 27&  22.13$\pm$0.03 \\
  10             &   14 23 27.0& 0.09 &             &   1.36 & 13.57& 24.2 & 23.02   & 22.67 &  13.57 &  40.79 & 16    &  21.34$\pm$0.03 \\
KDG52$^3$        &   08 23 56.0& 27.75&     -1.90   &   1.30 & 16.67& 25.8 & 0.95$^a$& 24.78 &  17.23 &  20.62 & 48    &  24.42$\pm$0.02 \\
  10             &   71 01 46.0& 0.07 &  0.09,0.15  &   1.06 & 15.72& 24.7 &         & 24.10 &  15.72 &  31.67 & 29    &  23.69$\pm$0.01 \\
DDO52$^{S,1}$    &   08 28 28.5& 30.03&             &   2.00 & 14.86& 27.3 &  0.89   & 24.14 &  15.43 &  34.06 & 30&  23.07$\pm$0.02 \\
  10             &   41 51 24.0& 0.12 &             &   2.20 & 14.02& 26.7 & 24.03   & 23.70 &  14.15 &  49.50 & 18    &  22.11$\pm$0.03 \\
\end{tabular}
\end{center}
\end{table*}
\setcounter{table}{0}
\begin{table*}
\begin{center}
\scriptsize
\caption{Continued.}
\begin{tabular}{lrrlccclccclc}\\ \hline \hline
Name              &  RA(2000.0)& $\mu_0$& \feh,$\sigma_{VI}$      &   Diam   & V$_L$& $SBV_{L}$& $(V\!-\!I)_e$& $SBV_{25}$ & $V_{25}$ & $R_{V,25}$& $R_e$ & $SBV_{C}$  \\
T          & DEC(2000.) & $A_v$   & $ \sigma$\feh  & $D_{lim}$& $I_L$& $SBI_{L}$ & $SBV_{e}$  & $SBI_{25}$ & $I_{25}$ & $R_{I,25}$& h   & $SBI_{C}$  \\
\hline
DDO53$^3$        &   08 34 06.5& 27.76& -1.93      &   1.60 &      & 24.6 & 0.63$^a$&       &        &        & 73&  23.75$\pm$0.02 \\
  10             &   66 10 45.0& 0.12 & 0.06,0.15  &   1.00 &      & 23.7 &         &       &        &        & 44    &  23.29$\pm$0.01 \\
D564-08$^{S,1}$  &   09 02 54.0& 29.61&            &   0.70 & 16.58& 27.0 &  0.86   & 24.31 &  17.46 &  13.86 & 18&  23.61$\pm$0.03 \\
  10             &   20 04 31.0& 0.10 &            &   1.02 & 15.85& 25.6 & 24.51   & 23.89 &  16.08 &  22.57 & 11    &  22.81$\pm$0.04 \\
D634-03$^1$      &   09 08 53.5& 29.90&            &   0.40 & 17.21& 26.8 &  0.86   & 24.31 &  18.41 &   9.00 & 15&  23.82$\pm$0.01 \\
  10             &   14 34 55.0& 0.13 &            &   0.91 & 16.34& 26.1 &  24.77  & 23.99 &  16.65 &  18.25 &  9    &  22.83$\pm$0.01 \\
D565-06$^{S,1}$  &   09 19 29.4& 29.79&            &   0.70 & 16.63& 27.3 &  0.90   & 23.96 &  17.28 &  13.46 & 14&  23.13$\pm$0.04 \\
  10             &   21 36 12.0& 0.13 &            &   0.96 & 15.78& 25.5 &  24.03  & 23.70 &  15.96 &  22.18 &  8    &  22.29$\pm$0.04 \\
FM1$^3$          &   09 45 10.0& 27.67& -1.20      &   0.90 & 16.83& 25.6 & 0.80    & 24.83 &  18.39 &  11.35 & 32    &  24.63$\pm$0.01 \\
  -3             &   68 45 54.0& 0.26 & 0.05,0.11  &   1.00 & 16.06& 25.5 &         & 24.17 &  16.19 &  25.50 & 25    &  23.50$\pm$0.02 \\
KK77$^3$         &   09 50 10.0& 27.71&            &   2.40 &      & 25.7 & 1.30$^a$& 24.79 &  19.93 &   5.48 & 31    &  24.62$\pm$0.01 \\
  -3             &   67 30 24.0& 0.48 &            &   0.66 &      & 24.1 &         & 23.87 &  16.37 &  19.92 & 18    &  23.55$\pm$0.01 \\
KDG61$^{S,3}$    &   09 57 02.7& 27.78& -1.51      &   2.40 & 14.98& 27.0 &  1.04   & 24.28 &  16.06 &  28.25 & 48&  24.00$\pm$0.06 \\
  -1             &   68 35 30.0& 0.24 & 0.04,0.13  &   2.69 & 13.92& 26.4 & 24.59   & 23.66 &  14.18 &  56.11 & 29    &  22.96$\pm$0.09 \\
KKH57$^{S,3}$    &   10 00 16.0& 27.97&  -1.37     &   0.60 & 17.73& 26.4 &  0.81   & 24.76 &  19.27 &   7.13 & 19&  24.38$\pm$0.03 \\
  -3             &   63 11 06.0& 0.08 & 0.03,0.12  &   0.74 & 16.90& 25.4 & 25.31   & 24.29 &  17.49 &  13.46 & 11    &  23.62$\pm$0.03 \\
ANTLIA$^2$       &   10 04 04.0& 25.49& -1.81      &   2.00 & 14.70& 25.3 & 0.60$^a$& 24.45 &  14.92 &  49.05 & 66&  23.94$\pm$0.01 \\
  10             &  -27 19 55.0& 0.26 &  0.04,0.15 &   1.91 & 14.14& 25.0 &         & 24.03 &  14.16 &  56.50 & 43    &  23.26$\pm$0.01 \\
DDO71$^3$        &   10 05 07.3& 27.72&  -1.17     &   1.70 & 16.46& 24.8 & 0.98$^a$& 24.13 &  16.46 &  21.41 & 59    &  23.80$\pm$0.01 \\
  -3             &   66 33 18.0& 0.32 &  0.04,0.10 &   0.71 & 15.48& 23.7 &         &       &        &        & 35    &  22.81$\pm$0.01 \\
KK84$^3$         &   10 05 34.4& 29.93&            &   1.30 &      & 25.5 &  1.30   & 23.85 &  16.19 &  22.91 & 19&  22.72$\pm$0.02 \\
  -3             &  -07 44 57.0& 0.16 &            &   0.88 &      & 24.0 & 23.62   & 22.78 &        &        & 12    &  21.66$\pm$0.02 \\
KDG64$^3$        &   10 07 01.9& 27.84&  -1.10     &   1.90 &      & 25.2 & 1.06$^a$& 24.01 &  15.89 &  28.09 & 28&  23.07$\pm$0.01 \\
  -3             &   67 49 39.0& 0.18 & 0.04,0.11  &   1.00 &      & 23.9 & 23.98   &       &        &        & 17    &  22.11$\pm$0.01 \\
HS117$^1$        &   10 21 25.2& 27.97& -1.87,0.08 &   1.50 & 16.16& 28.1 &  1.10   & 24.75 &  17.97 &  13.15 & 29&  24.32$\pm$0.01 \\
  10             &   71 06 58.0& 0.38 & 0.03,0.15  &   2.08 & 15.15& 27.1 & 25.20   & 24.13 &  15.76 &  28.35 & 18    &  23.30$\pm$0.01 \\
BK6N$^3$         &   10 34 31.9& 27.93&  -1.36     &   1.10 &      & 25.8 & 1.20$^a$& 24.90 &  18.26 &  13.15 & 40    &  24.61$\pm$0.02 \\
  -3             &   66 00 42.0& 0.04 & 0.08,0.12  &   0.76 &      & 24.4 &         & 23.95 &  16.13 &  22.81 & 28    &  23.55$\pm$0.02 \\
E215-09$^1$      &   10 57 30.2& 28.60&  -1.86     &   2.00 &      & 26.0 & 1.26$^a$& 24.53 &  15.53 &  36.40 & 60    &  24.04$\pm$0.01 \\
10               &  -48 10 44.0& 0.73 & 0.02,0.15  &   1.63 &      & 24.9 &         & 23.64 &  13.85 &  49.05 & 38    &  22.89$\pm$0.01 \\
UGC6541$^{S,3}$  &   11 33 29.1& 27.95&  -1.57     &   1.40 & 14.03& 26.0 &  0.46   & 23.22 &  14.17 &  35.07 & 18&  21.25$\pm$0.06 \\
  10             &   49 14 17.0& 0.06 & 0.05,0.15  &   1.51 & 13.54& 24.9 & 22.59   & 22.68 &  13.54 &  45.39 & 11    &  20.75$\pm$0.03 \\
NGC3741$^{S,3}$  &   11 36 06.4& 27.41&   -1.74    &   2.00 & 14.04& 27.3 &  0.61   & 23.17 &  14.17 &  39.20 & 19&  21.28$\pm$0.02 \\
  10             &   45 17 07.0& 0.09 & 0.06,0.15  &   2.01 & 13.42& 26.1 & 22.16   & 22.87 &  13.50 &  45.94 & 12    &  20.87$\pm$0.02 \\
E320-14$^1$      &   11 37 53.4& 28.76&   -2.28    &   1.40 & 15.49& 26.5 &  0.93   & 23.79 &  15.71 &  25.65 & 17&  22.37$\pm$0.01 \\
  10             &  -39 13 14.0& 0.47 & 0.08,0.18  &   1.24 & 14.58& 25.6 & 23.34   & 23.24 &  14.64 &  32.00 & 10    &  21.52$\pm$0.01 \\
KK109$^{3,S}$    &   11 47 11.2& 28.27&            &   0.60 & 17.77& 25.9 &  0.71   & 24.26 &  18.32 &   9.56 & 12&  23.48$\pm$0.02 \\
  10             &   43 40 19.0&  0.06&            &   0.60 & 17.06& 25.2 &  24.41  & 23.80 &  17.30 &  12.65 &  7    &  22.76$\pm$0.02 \\
DDO99$^3$        &   11 50 53.0& 27.11&   -2.13    &   4.10 &      & 24.1 & 0.68$^a$&       &        &        & 35    &  22.70$\pm$0.01 \\
  10             &   38 52 50.0& 0.09 &  0.04,0.18 &   1.00 &      & 23.4 &         &       &        &        & 21    &  22.06$\pm$0.01 \\
E379-07$^3$      &   11 54 43.0& 28.59&            &   1.10 & 16.41& 25.8 & 1.00$^a$& 24.21 &  16.82 &  20.82 & 27&  23.44$\pm$0.01 \\
  10             &  -33 33 29.0& 0.25 &            &   1.03 & 15.39& 24.4 & 24.35   & 23.58 &  15.39 &  30.78 & 17    &  22.70$\pm$0.01 \\
NGC4068$^{S,1}$  &   12 04 02.4& 28.16&            &   3.20 & 12.56& 24.6 &  0.86   & 23.15 &  12.70 &  82.86 & 45&  21.72$\pm$0.03 \\
10               &   52 35 19.0& 0.07 &            &   3.27 & 11.97& 25.9 & 22.55   & 22.41 &  11.97 &  98.13 & 27    &  21.06$\pm$0.02 \\
NGC4163$^{S,1}$  &   12 12 08.9& 27.78&   -1.65    &   1.90 & 12.96& 27.4 &  0.80   & 23.04 &  13.15 &  57.52 & 27&  21.09$\pm$0.04 \\
  10             &   36 10 10.0& 0.06 &  0.01,0.14 &   2.70 & 12.16& 25.7 & 22.24   & 22.19 &  12.17 &  78.55 & 16    &  20.25$\pm$0.03 \\
E321-014$^3$     &   12 13 49.6& 27.52&  -2.28     &   1.40 & 16.06& 23.8 & 0.92$^a$&       &        &        & 33    &  22.80$\pm$0.02 \\
  10             &  -38 13 53.0& 0.31 &  0.08,0.18 &   0.60 & 15.14& 22.9 &         &       &        &        & 20    &  21.97$\pm$0.02 \\
UGC7242$^1$      &   12 14 07.4& 28.58&  -1.49     &   1.90 & 13.87& 26.4 &  0.98   & 23.03 &  13.92 &  41.00 & 20&  21.06$\pm$0.01 \\
  10             &   66 05 32.0& 0.06 &  0.02,0.13 &   1.64 & 12.97& 24.9 & 22.04   & 22.50 &  12.97 &  49.25 & 12    &  20.44$\pm$0.01 \\
DDO113$^{S,3}$   &   12 14 57.9& 27.28&  -1.99     &   1.50 & 15.41& 27.0 &  1.01   & 24.64 &  16.42 &  25.74 & 42&  24.05$\pm$0.03 \\
  10             &   36 13 08.0& 0.07 &  0.05,0.16 &   2.14 & 14.39& 25.5 & 24.93   & 24.11 &  14.61 &  48.71 & 25    &  23.12$\pm$0.02 \\
UGC7298$^{S,3}$  &   12 16 28.6& 28.12&            &   1.10 &      & 25.7 &  0.76   & 23.63 &  15.87 &  21.38 & 17&  22.40$\pm$0.05 \\
  10             &   52 13 38.0& 0.08 &            &   0.86 &      & 24.7 & 23.28   & 23.28 &  15.15 &  25.74 & 10    &  22.02$\pm$0.04 \\
UGC7369$^{S,1}$  &   12 19 38.7& 30.52&            &   1.00 & 14.22& 26.6 &  1.07   & 22.71 &  14.34 &  34.62 & 15&  21.02$\pm$0.02 \\
  10             &   29 52 59.0& 0.06 &            &   1.51 & 13.14& 25.3 &  21.92  & 21.75 &  13.17 &  41.96 &  9    &  19.84$\pm$0.02 \\
DDO125$^{S,3}$   &   12 27 41.8& 27.02&  -1.73     &   4.30 & 12.51& 26.4 &  0.93   & 23.54 &  12.69 &  91.87 & 60    &  22.11$\pm$0.01 \\
  10             &   43 29 38.0& 0.07 &  0.02,0.15 &   4.24 & 11.66& 25.1 & 23.03   & 23.18 &  11.66 &  126.3 & 37    &  21.42$\pm$0.01 \\
UGC7605$^{S,3}$  &   12 28 39.0& 28.23&            &   1.10 & 14.46& 26.4 &  0.56   & 23.49 &  14.66 &  36.04 & 22&  21.95$\pm$0.02 \\
  10             &   35 43 05.0& 0.05 &            &   1.76 & 13.88& 25.6 & 22.83   & 23.20 &  13.95 &  43.96 & 13    &  21.46$\pm$0.02 \\
E381-018$^1$     &   12 44 42.7& 28.63&   -2.06    &   1.20 & 15.28& 27.7 &  0.66   & 23.36 &  15.55 &  22.90 & 14&  21.91$\pm$0.01 \\
  10             &  -35 58 00.0& 0.21 &  0.04,0.16 &   1.48 & 14.55& 26.5 &  22.78  & 23.20 &  14.67 &  31.15 &  9    &  21.32$\pm$0.01 \\
E381-20$^1$      &   12 46 00.4& 28.69&   -2.34    &   3.00 &      & 25.2 &  0.53   & 23.74 &  14.06 &  53.70 & 39&  22.55$\pm$0.01 \\
  10             &  -33 50 17.0& 0.21 &  0.03,0.18 &   2.02 &      & 24.9 & 23.43   & 23.32 &  13.37 &  60.45 & 23    &  21.87$\pm$0.01 \\
HIPASS1247-77$^1$& 12 47 32.6  &29.41 &            &   0.80 & 17.49& 27.7 &  1.29   & 24.62 &  18.59 &   8.55 & 13&  23.93$\pm$0.02 \\
  10             &  -77 35 01.0& 2.48 &            &   1.00 & 16.25& 26.5 &  24.88  & 23.94 &  16.55 &  17.50 &  8    &  22.70$\pm$0.01 \\
KK166$^{2,3}$    &   12 49 13.3& 28.38&            &   0.70 & 17.50& 25.9 & 1.20$^a$& 24.92 &  18.52 &  12.15 & 38    &  24.62$\pm$0.02 \\
  -3             &   35 36 45.0& 0.05 &            &   0.81 & 16.22& 24.5 &         & 23.98 &  16.22 &  24.40 & 26    &  23.51$\pm$0.02 \\
E443-09$^1$      &   12 54 53.6& 28.81&  -2.43     &   0.80 & 16.73& 27.0 &  0.79   & 24.35 &  17.65 &  12.60 & 20&  23.71$\pm$0.01 \\
  10             &  -28 20 27.0& 0.22 &  0.08,0.18 &   1.09 & 16.01& 25.9 & 24.64   & 24.17 &  16.47 &  20.80 & 12    &  23.32$\pm$0.01 \\
E269-37$^3$      &   13 03 33.6& 27.71&            &   0.80 & 16.10& 24.8 & 0.80$^a$& 23.95 &  16.10 &  26.19 & 25    &  23.04$\pm$0.02 \\
  -3             &  -46 35 03.0& 0.44 &            &   0.87 & 15.32& 24.2 &         & 23.16 &  15.32 &  26.19 & 15    &  22.10$\pm$0.01 \\
KK182$^1$        &   13 05 02.9& 28.85&  -2.74     &   1.00 & 15.92& 27.6 &  0.93   & 23.97 &  16.18 &  22.20 & 15&  22.67$\pm$0.02 \\
  10             &  -40 04 58.0& 0.34 & 0.08,0.19  &   1.48 & 15.04& 26.4 &  23.62  & 23.41 &  15.27 &  26.05 &  9    &  22.05$\pm$0.01 \\
UGC8215$^{1,S}$  &   13 08 03.6& 28.74&  -1.67,0.11&   1.00 & 15.61& 26.4 &  0.62   & 23.75 &  15.86 &  23.10 & 16&  22.43$\pm$0.01 \\
  10             &   46 49 41.0& 0.04 & 0.03,0.14  &   1.18 & 15.00& 25.9 &  23.36  & 23.40 &  15.12 &  27.60 & 10    &  21.91$\pm$0.01 \\
\end{tabular}
\end{center}
\end{table*}
\setcounter{table}{0}
\begin{table*}
\begin{center}
\scriptsize
\caption{Continued.}
\begin{tabular}{lrrlccclccclc}\\ \hline \hline
Name              &  RA(2000.0)& $\mu_0$& \feh,$\sigma_{VI}$      &   Diam   & V$_L$& $SBV_{L}$& $(V\!-\!I)_e$& $SBV_{25}$ & $V_{25}$ & $R_{V,25}$ & $R_e$ & $SBV_{C}$  \\
T          & DEC(2000.) & $A_v$   & $ \sigma$\feh  & $D_{lim}$& $I_L$& $SBI_{L}$ & $SBV_{e}$  & $SBI_{25}$ & $I_{25}$ & $R_{I,25}$& h   & $SBI_{C}$  \\
\hline
E269-58$^1$      &   13 10 32.9& 27.84&  -1.43,0.23&   3.00 & 12.19& 25.3 &   1.13   & 22.81 & 12.21 & 82.00&  36    & 20.70$\pm$0.03 \\
10               &  -46 59 27.0& 0.35 &  0.01,0.12 &   2.90 & 11.08& 24.0 &  21.63   & 21.83 & 11.08 & 86.95&  21    & 19.73$\pm$0.03 \\
KK189$^1$        &   13 12 45.0& 28.12& -1.74,0.08 &   0.60 & 16.98& 25.8 &  1.20$^a$& 24.03 & 17.32 & 17.40&  20& 23.82$\pm$0.04 \\
  -3             &  -41 49 55.0& 0.38 & 0.04,0.14  &   0.81 & 15.78& 24.5 &  24.19   & 23.30 & 15.95 & 21.23&  13    & 22.97$\pm$0.06 \\
E269-66$^1$      &   13 13 09.2& 27.82& -1.22,0.19 &   1.40 & 13.74& 26.2 &   1.06   & 23.81 & 13.94 & 57.95&  40    & 22.52$\pm$0.01 \\
-1               &  -44 53 24.0& 0.31 &  0.01,0.11 &   2.57 & 12.65& 25.5 &  23.44   & 23.07 & 12.66 & 74.20&  24    & 21.38$\pm$0.01 \\
DDO167$^{S,3}$   &   13 13 22.8& 28.11&            &   1.10 & 15.08& 27.7 &   0.48   & 23.77 & 15.40 & 28.51&  21& 22.57$\pm$0.04 \\
  10             &   46 19 11.0& 0.03 &            &   1.50 & 14.63& 26.5 &  23.53   & 23.45 & 14.71 & 34.45&  12    & 22.03$\pm$0.07 \\
KK196$^1$        &   13 21 47.1& 27.99&  -1.96,0.10&   1.30 & 15.47& 27.0 &   1.09   & 23.94 & 16.02 & 23.15&  21& 22.84$\pm$0.01 \\
  10             &  -45 03 48.0& 0.28 &  0.03,0.15 &   1.50 & 14.41& 25.6 &  23.77   & 23.54 & 14.53 & 36.90&  13    & 22.07$\pm$0.01 \\
KK197$^1$        &   13 22 01.8& 27.87&  -1.19,0.24&   0.90 & 15.31& 26.1 &   1.27   & 24.15 & 16.50 & 19.69&  38& 23.78$\pm$0.03 \\
  -3             &  -42 32 08.0& 0.51 &  0.02,0.10 &   1.70 & 14.04& 24.7 &  24.26   & 23.15 & 14.14 & 45.20&  23    & 22.55$\pm$0.03 \\
KKs55$^1$        &   13 22 12.4& 27.98&  -1.81,0.12&   0.80 & 15.53& 27.8 &   1.12   & 25.30 & 16.05 & 71.65&  50& 24.88$\pm$0.01 \\
  -3             &  -42 43 51.0& 0.48 &  0.04,0.15 &   2.39 & 14.40& 26.4 &  25.74   & 24.30 & 14.91 & 42.55&  31    & 23.53$\pm$0.01 \\
KK200$^3$        &   13 24 36.0& 28.33&            &   1.30 &      & 26.0 &   1.01   & 23.72 & 15.96 & 25.40&  19    & 22.42$\pm$0.02 \\
   9             &  -30 58 20.0& 0.23 &            &   1.09 &      & 24.7 &  23.31   & 23.18 & 14.88 & 32.77&  11    & 21.70$\pm$0.01 \\
I4247$^1$        &   13 26 44.4& 28.48& -2.37,0.18 &   1.30 & 13.97& 26.7 &   0.68   & 22.90 & 14.09 & 35.55&  16    & 20.88$\pm$0.01 \\
10               &  -30 21 45.0& 0.21 & 0.03,0.18  &   1.90 & 13.24& 26.3 &   21.73  & 22.56 & 13.29 & 42.95&  10    & 20.13$\pm$0.02   \\
UGC8508$^{S,3}$  &   13 30 44.4& 27.04& -1.91      &   1.70 & 13.63& 27.1 &   0.72   & 23.25 & 13.77 & 48.31&  25& 21.50$\pm$0.02 \\
  10             &   54 54 36.0& 0.05 & 0.04,0.15  &   2.42 & 12.94& 26.4 &  22.43   & 22.96 & 12.99 & 60.59&  15    & 20.90$\pm$0.02 \\
E444-78$^1$      &   13 36 30.8& 28.56& -1.75,0.13 &   1.20 & 14.78& 27.5 &   0.85   & 23.79 & 15.22 & 31.95&  26    & 22.68$\pm$0.01 \\
10               &  -29 14 11.0& 0.18 & 0.02,0.14  &   2.51 & 13.89& 26.9 &  23.56   & 23.44 & 14.10 & 45.75&  15    & 21.87$\pm$0.01 \\
E444-84$^3$      &   13 37 20.2& 28.32&            &   1.30 & 14.98& 24.8 &   0.77   & 23.48 & 14.98 & 35.56&  31    & 22.40$\pm$0.02 \\
  10             &  -28 02 46.0& 0.23 &            &   1.19 & 14.26& 24.3 &  23.28   & 22.78 & 14.26 & 35.56&  18    & 21.76$\pm$0.02 \\
NGC5237$^1$      &   13 37 38.9& 27.61&            &   1.90 & 12.36& 25.6 &   1.01   & 22.75 & 12.38 & 70.70&  25    & 20.18$\pm$0.01 \\
-3               &  -42 50 51.0& 0.32 &            &   2.59 & 11.31& 24.6 &  21.02   & 21.97 & 11.31 & 77.80&  15    & 19.20$\pm$0.01 \\
UGC8638$^{1,S}$  &   13 39 19.4& 26.81& -1.64,0.16 &   1.20 & 13.85& 26.6 &   0.72   & 23.47 & 14.07 & 44.30&  29& 21.98$\pm$0.01 \\
  10             &   24 46 33.0& 0.04 & 0.02,0.14  &   2.57 & 13.11& 25.7 &  22.88   & 23.17 & 13.22 & 59.10&  17    & 21.33$\pm$0.01 \\
UGC8651$^{2,3}$  &   13 39 53.8& 27.39& -1.92, 0.11&   2.30 &      & 24.6 &   0.64   &       &       &      &  41& 22.70$\pm$0.01 \\
  10             &   40 44 21.0& 0.02 & 0.02,0.15  &   1.57 &      & 24.1 &  23.62   & 23.08 &       &      &  24    & 22.02$\pm$0.01 \\
I4316$^3$        &   13 40 18.1& 28.22&            &   1.60 &      & 23.8 &   0.98   &       &       &      &  21& 21.68$\pm$0.03 \\
  10             &  -28 53 40.0& 0.18 &            &   0.81 &      & 22.9 &  22.51   &       &       &      &  12    & 20.98$\pm$0.02 \\
KKs57$^1$        &   13 41 38.1& 27.89& -1.63, 0.08&   0.80 & 17.28& 27.0 &   1.60   & 24.67 & 18.77 &  8.30&  17    & 24.23$\pm$0.01 \\
-3               &  -42 34 55.0& 0.30 & 0.06,0.14  &   0.95 & 15.82& 25.9 &  25.10   & 23.97 & 16.11 & 18.80&  10    & 23.04$\pm$0.01 \\
KK211$^3$        &   13 42 05.6& 27.77&   -1.26    &   0.80 & 15.69& 25.2 &   1.14   & 23.91 & 15.75 & 30.68&  28& 22.91$\pm$0.03 \\
  -5             &  -45 12 18.0& 0.37 & 0.05,0.12  &   1.09 & 14.56& 23.9 &  23.82   &       &       &      &  17    & 21.78$\pm$0.03 \\
KK213$^3$        &   13 43 35.8& 27.80&   -1.62    &   0.60 & 17.70& 25.6 &  1.20$^a$& 24.20 & 17.92 & 16.43&  18& 23.31$\pm$0.02 \\
  -3             &  -43 46 09.0& 0.32 & 0.09,0.14  &   0.68 & 16.50& 23.8 &  24.22   &       &       &      &  11    & 22.27$\pm$0.02 \\
E325-11$^3$      &   13 45 00.8& 27.66&   -2.18    &   2.70 & 15.73& 24.2 &  0.96$^a$&       &       &      &  40    & 23.16$\pm$0.02 \\
  10             &  -41 51 32.0& 0.24 & 0.05,0.17  &   0.88 & 14.77& 23.2 &          &       &       &      &  24    & 22.10$\pm$0.03 \\
KK217$^3$        &   13 46 17.2& 27.92&   -1.37    &   0.60 & 16.59& 25.9 &   1.23   & 23.70 & 16.75 & 22.71&  17& 22.45$\pm$0.01 \\
  -3             &  -45 41 05.0& 0.40 & 0.07,0.13  &   1.00 & 15.46& 24.5 &  23.36   & 22.97 & 15.46 & 29.88&  10    & 21.47$\pm$0.01 \\
CenN$^1$         &   13 48 09.2& 27.82& -1.74,0.10 &   0.90 & 16.42& 27.2 &   1.37   & 24.50 & 17.20 & 16.75&  25    & 23.90$\pm$0.01 \\
-3               &  -47 33 54.0& 0.47 &  0.03,0.14 &   1.21 & 15.12& 25.6 &  24.78   & 23.89 & 15.16 & 34.20&  15    & 22.80$\pm$0.01 \\
HIPASS1348-37$^1$&   13 48 33.9& 28.78&    -2.52   &   0.50 & 17.00& 25.9 &  0.60$^a$& 24.77 & 18.55 & 10.10&  33    & 24.53$\pm$0.01 \\
10               &  -37 58 03.0& 0.26 &  0.06,0.19 &   0.91 & 16.45& 25.8 &          & 24.50 & 16.65 & 22.45&  20    & 23.93$\pm$0.01 \\
KK221$^3$        &   13 48 46.4& 28.00&            &   1.50 &      & 25.5 &  1.30$^a$& 24.82 & 17.51 & 17.83&        & 24.67$\pm$0.02 \\
  -3             &  -46 59 49.0& 0.46 &            &   0.88 &      & 24.2 &          & 23.64 & 15.46 & 26.39&        & 23.40$\pm$0.02 \\
UGC8760$^{S,2}$  &   13 50 51.1& 28.54& -2.00,0.16 &   2.20 & 14.14& 27.9 &   0.82   & 23.79 & 14.36 & 49.50&  31& 22.32$\pm$0.03 \\
  10             &   38 01 16.0& 0.05 & 0.03,0.16  &   3.00 & 13.40& 26.3 &  23.30   & 23.40 & 13.51 & 60.59&  18    & 21.74$\pm$0.02 \\
HIPASS1351-47$^1$&   13 51 22.0& 28.63&            &   0.50 & 16.51& 25.9 &  0.70$^a$& 24.65 & 17.70 & 14.20&  33    & 24.28$\pm$0.01 \\
10               &  -47 00 00.0& 0.48 &            &   1.03 & 15.83& 25.4 &          & 24.26 & 16.15 & 23.85&  20    & 23.58$\pm$0.01 \\
KKH86$^{S,3}$    &   13 54 33.6& 27.08&  -2.33     &   0.70 & 16.55& 26.7 &   0.91   & 24.15 & 17.11 & 15.44&  17    & 23.24$\pm$0.03 \\
  10             &   04 14 35.0& 0.09 & 0.11,0.18  &   0.82 & 15.86& 25.7 &  24.17   & 23.75 & 15.91 & 22.57&  10    & 22.60$\pm$0.02 \\
UGC8833$^{2,S}$  &   13 54 48.7& 27.52& -2.03,0.07 &   0.90 & 14.86& 26.4 &   0.62   & 23.66 & 15.11 & 30.25&  22& 22.39$\pm$0.01 \\
  10             &   35 50 15.0& 0.04 & 0.03,0.16  &   1.54 & 14.23& 25.8 &  23.30   & 23.39 & 14.31 & 39.45&  14    & 21.88$\pm$0.01 \\
E384-016$^1$     &   13 57 01.6& 28.26&            &   1.30 & 14.37& 26.7 &   1.00   & 23.37 & 14.65 & 34.00&  21    & 21.84$\pm$0.01 \\
10               &  -35 20 02.0& 0.25 &            &   2.04 & 13.39& 26.0 &  22.72   & 22.91 & 13.48 & 47.75&  13    & 20.89$\pm$0.01 \\
KK230$^1$        &   14 07 10.7& 26.34&  -2.02     &   0.60 & 16.31& 28.2 &   0.58   & 24.44 & 17.25 & 14.95&  21& 23.78$\pm$0.01 \\
  10             &   35 03 37.0&  0.05& 0.09,0.16  &   1.73 & 15.68& 27.4 &  24.70   & 24.19 & 16.25 & 23.10&  13    & 23.28$\pm$0.01 \\
UGC9128$^1$      &   14 15 56.5& 26.75& -2.33,0.08 &   1.70 & 14.38& 24.9 &   0.75   & 23.37 & 14.38 & 46.83&  38& 22.60$\pm$0.03 \\
  10             &   23 03 19.0& 0.08 & 0.06,0.18  &   1.56 & 13.75& 24.1 &  23.10   & 22.78 & 13.75 & 46.83&  23    & 22.08$\pm$0.03 \\
DDO190$^{S,3}$   &   14 24 43.5& 27.23&   -2.09    &   1.80 & 13.13& 24.1 &   0.71   & 22.89 & 13.13 & 51.53&  38    & 21.45$\pm$0.04 \\
  10             &   44 31 33.0& 0.04 & 0.04,0.17  &   1.85 & 12.39& 23.6 &  22.49   & 22.29 & 12.39 & 55.58&  23    & 20.85$\pm$0.04 \\
E223-09$^1$      &   15 01 08.5& 28.95&            &   2.60 & 12.83& 24.2 &   1.00   &       &       &      &  64& 22.17$\pm$0.01 \\
  10             &  -48 17 33.0& 0.86 &            &   2.32 & 11.77& 23.2 &  23.02   &       &       &      &  38    & 21.03$\pm$0.01 \\
E137-18$^1$      &   16 20 59.3& 29.01&            &   3.20 &      & 23.5 &   1.33   &       &       &      &  34& 20.92$\pm$0.07 \\
   9             &  -60 29 15.0& 0.81 &            &   1.96 &      & 22.1 &  21.97   &       &       &      &  21    & 19.67$\pm$0.08 \\
IC4662$^1$       &   17 47 06.3& 26.90& -1.34,0.23 &   2.80 & 11.13& 25.5 &   0.57   & 22.28 & 11.14 & 88.80&  28& 19.24$\pm$0.01 \\
   9             &  -64 38 25.0& 0.23 &  0.01,0.12 &   3.17 & 10.47& 24.7 &  20.04   & 21.66 & 10.47 & 95.00&  17    & 18.74$\pm$0.01 \\
KK246$^1$        &   20 03 57.4& 29.78&            &   1.20 & 16.50& 27.4 &   1.37   & 24.21 & 17.37 & 14.15&  17& 23.40$\pm$0.01 \\
  10             &  -31 40 54.0& 0.98 &            &   1.42 & 15.17& 26.1 &  24.33   & 23.61 & 15.39 & 27.30&  10    & 22.27$\pm$0.01 \\
UGCA438$^3$      &   23 26 27.5& 26.59&    -1.68   &   1.50 &      & 23.6 &  0.81$^a$&       &       &      &  50& 22.43$\pm$0.02 \\
  10             &  -32 23 26.0& 0.05 & 0.04,0.15  &   0.97 &      & 22.8 &          &       &       &      &  30    & 21.79$\pm$0.03 \\
KKH98$^3$        &   23 45 34.0& 26.95&   -1.94    &   1.10 & 15.16& 24.2 &  0.60    &       &       &      &  38& 22.81$\pm$0.01 \\
  10             &   38 43 04.0& 0.41 & 0.09,0.16  &   1.07 & 14.61& 24.1 &          & 22.91 &       &      &  23    & 21.98$\pm$0.02 \\
\hline \hline
\end{tabular}
\end{center}
\end{table*}
\begin{table*}
\begin{center}
\scriptsize
\caption{Fundamental photometric parameters of galaxies from Table~1 measured on the SDSS images in the g and r bands.
The data were transformed into B and R bands of the Johnson-Cousins system using the transformations by Jordi et al. (2006).
Columns contain the following data:
{\bf (1)} Galaxy Name.
{\bf (2)} equatorial coordinates (J2000),
{\bf (3)} projected major axis from CNG and limiting diameter of our photometry,
{\bf (4)} integrated apparent B and R magnitudes,
{\bf (5)} limiting B and R surface brightnesses (SB) taken at the distance D$_{lim}$/2 from centers of galaxies,
{\bf (6)} effective $(B\!-\!R)$ color and effective SB in the R-band,
{\bf (7)} mean SB within the isophote of 25$^m\!/\sq \arcsec$ in the B and R bands,
{\bf (8)} apparent B and R magnitudes integrated within the isophote level 25 $^m\!/\sq \arcsec$,
{\bf (9)} projected major axis at the isophote level 25$^m\!/\sq \arcsec$ in the B and R bands,
{\bf (10)} effective radius and model exponential scale length,
{\bf (11)} central SB in B and R bands and the corresponding errors.
The data were not corrected for Galactic extinction.}
\begin{tabular}{lrcclcccccc}\\ \hline \hline
 Name   &  RA(2000)& diam     & $B_L$& $SBB_L$& $(B-R)_e$& $SBB_{25}$& $B_{25}$& $R_{B,25}$& $R_e$& $SBB_C$ \\
	& DEC(2000)& D$_{lim}$& $R_L$& $SBR_L$&  $SBR_e$& $SBR_{25}$& $R_{25}$& $R_{R,25}$& h    & $SBI_C$ \\
	   &            & $\arcmin$& $^{m}$& $^{m}\!/\sq \arcsec$& $^{m},^{m}\!/\sq \arcsec$& $^{m}\!/\sq \arcsec$& $^{m}$& $\arcsec$& $\arcsec$& $^{m}\!/\sq \arcsec$\\
(1)        &    (2)& (3) & (4) & (5) & (6) & (7) & (8) & (9) & 10 & 11 \\
\hline
UGC3974 &  07 41 55.0& 3.10 & 14.61 & 25.35 &  0.77& 24.24& 15.06& 43.56&   54& 23.55$\pm$0.03 \\
	&  16 48 02.0& 1.97 & 13.88 & 24.66 & 23.67& 23.75& 13.88& 59.00&   32& 22.74$\pm$0.03 \\
 KK65   &  07 42 31.2& 0.90 & 15.46 & 25.71 &  0.87& 23.61& 15.76& 22.97&   17& 22.38$\pm$0.03 \\
	&  16 33 40.0& 1.10 & 14.64 & 24.86 & 22.42& 23.27& 14.64& 32.87&   10& 21.50$\pm$0.02 \\
UGC4115 &  07 57 01.8& 1.80 & 14.84 & 25.48 &  0.84& 23.83& 14.99& 35.64&   27& 22.61$\pm$0.03 \\
	&  14 23 27.0& 1.36 & 14.06 & 24.62 & 22.68& 23.17& 14.06& 40.79&   16& 21.79$\pm$0.03 \\
 DDO52  &  08 28 28.5& 2.00 & 15.36 & 27.88 &  0.88& 24.36& 16.43& 23.76&   31& 23.63$\pm$0.03 \\
	&  41 51 24.0& 2.20 & 14.52 & 26.88 & 23.62& 23.90& 14.83& 41.58&   19& 22.68$\pm$0.02 \\
 D564-08&  09 02 54.0& 0.70 & 17.16 & 27.47 &  0.97& 24.66& 18.97&  7.92&   18& 24.11$\pm$0.04 \\
	&  20 04 31.0& 1.02 & 16.30 & 26.13 & 24.15& 24.15& 16.73& 18.61&   11& 23.24$\pm$0.03 \\
 D565-06&  09 19 29.4& 0.70 & 17.13 & 27.63 &  0.87& 24.09& 18.17&  9.11&   14& 23.62$\pm$0.05 \\
	&  21 36 12.0& 0.96 & 16.33 & 26.28 & 23.64& 23.81& 16.75& 15.84&    8& 22.72$\pm$0.03 \\
 KDG61  &  09 57 02.7& 2.40 & 15.66 & 27.69 &  1.18& 24.83& 18.94&  8.25&   49& 24.81$\pm$0.07 \\
	&  68 35 30.0& 2.69 & 14.50 & 26.64 & 24.17& 24.03& 14.78& 56.11&   29& 23.44$\pm$0.06 \\
 KKH57  &  10 00 16.0& 0.60 & 18.34 & 27.43 &  1.00&  0.00&  0.00&  0.00&   19& 24.97$\pm$0.04 \\
	&  68 35 30.0& 0.74 & 17.29 & 26.29 & 24.94& 24.54& 18.21& 10.69&   11& 24.01$\pm$0.02 \\
UGC6541 &  11 33 29.1& 1.40 & 14.45 & 26.75 &  0.58& 23.63& 14.62& 30.00&   17& 21.50$\pm$0.07 \\
	&  49 14 17.0& 1.51 & 13.80 & 25.07 & 22.10& 22.79& 13.87& 38.85&   10& 20.92$\pm$0.03 \\
NGC3741 &  11 36 06.4& 2.00 & 14.44 & 27.52 &  0.64& 23.29& 14.67& 32.87&   19& 21.66$\pm$0.04 \\
	&  45 17 07.0& 2.01 & 13.77 & 26.52 & 21.97& 23.02& 13.87& 41.58&   11& 21.05$\pm$0.02 \\
 KK109  &  11 47 11.2& 0.60 & 18.20 & 27.64 &  0.53& 24.49& 19.18&  6.73&    9& 23.78$\pm$0.05 \\
	&  43 40 19.0& 0.54 & 17.78 & 26.43 & 24.15& 24.18& 18.23&  9.11&    5& 23.26$\pm$0.06 \\
NGC4068 &  12 04 02.4& 3.20 & 13.11 & 25.19 &  0.93& 23.42& 13.47& 68.00&   44& 22.09$\pm$0.03 \\
	&  52 35 19.0& 3.27 & 12.28 & 25.93 & 22.22& 22.80& 12.28& 98.13&   26& 21.46$\pm$0.02 \\
NGC4163 &  12 12 08.9& 1.90 & 13.55 & 28.20 &  0.89& 23.41& 13.81& 47.07&   26& 21.60$\pm$0.05 \\
	&  36 10 10.0& 2.70 & 12.63 & 26.42 & 21.67& 22.55& 12.67& 69.52&   16& 20.64$\pm$0.03 \\
 DDO113 &  12 14 57.9& 1.50 & 16.15 & 27.96 &  1.35& 24.96& 19.13&  8.71&   40& 24.63$\pm$0.04 \\
	&  36 13 08.0& 2.14 & 14.89 & 26.29 & 24.53& 24.35& 15.51& 34.85&   24& 23.64$\pm$0.02 \\
UGC7298 &  12 16 28.6& 1.10 & 16.05 & 26.06 &  0.70& 23.75& 16.28& 19.01&   16& 22.60$\pm$0.05 \\
	&  52 13 38.0& 0.86 & 15.51 & 25.12 & 23.15& 23.56& 15.55& 24.16&    9& 22.26$\pm$0.05 \\
UGC7369 &  12 19 38.7& 1.00 & 14.98 & 27.42 &  1.24& 23.22& 15.26& 25.92&   15& 21.81$\pm$0.03 \\
	&  29 52 59.0& 1.51 & 13.72 & 25.99 & 21.38& 22.26& 13.76& 39.68&    9& 20.45$\pm$0.02 \\
 DDO125 &  12 27 41.8& 4.30 & 13.03 & 26.81 &  0.92& 23.83& 13.35& 77.22&   59& 22.62$\pm$0.01 \\
	&  43 29 38.0& 4.24 & 12.14 & 25.83 & 22.71& 23.37& 12.22& 108.5&   35& 21.79$\pm$0.01 \\
UGC7605 &  12 28 39.0& 1.10 & 14.83 & 26.63 &  0.58& 23.64& 15.11& 31.28&   21& 22.25$\pm$0.03 \\
	&  35 43 05.0& 1.76 & 14.22 & 26.04 & 22.64& 23.38& 14.35& 39.60&   13& 21.73$\pm$0.01 \\
UGC8215 &  13 08 03.6& 1.00 & 16.03 & 27.21 &  0.80& 23.94& 16.36& 19.24&   16& 22.59$\pm$0.02 \\
	&  46 49 41.0& 1.06 & 15.39 & 26.14 & 23.18& 23.68& 15.48& 26.68&    9& 22.22$\pm$0.02 \\
 DDO167 &  13 13 22.8& 1.10 & 15.46 & 28.08 &  0.69& 23.97& 15.86& 25.34&   21& 22.87$\pm$0.04 \\
	&  46 19 11.0& 1.50 & 14.88 & 26.31 & 23.34& 23.67& 15.06& 32.08&   13& 22.41$\pm$0.05 \\
UGC8508 &  13 30 44.4& 1.70 & 14.07 & 27.56 &  0.73& 23.52& 14.27& 43.56&   25& 21.93$\pm$0.03 \\
	&  54 54 36.0& 2.42 & 13.33 & 26.53 & 22.17& 23.02& 13.44& 50.69&   15& 21.22$\pm$0.02 \\
UGC8638 &  13 39 19.4& 1.20 & 14.49 & 27.25 &  0.85& 23.49& 14.80& 33.66&   25& 22.24$\pm$0.04 \\
	&  24 46 33.0& 2.35 & 13.61 & 26.36 & 22.44& 23.29& 13.75& 47.92&   15& 21.53$\pm$0.03 \\
UGC8760 &  13 50 51.1& 2.20 & 14.58 & 28.68 &  0.85& 23.96& 15.00& 38.81&   30& 22.66$\pm$0.04 \\
	&  38 01 16.0& 3.00 & 13.80 & 26.95 & 23.06& 23.59& 14.02& 51.48&   18& 22.10$\pm$0.02 \\
 KKH86  &  13 54 33.6& 0.70 & 17.10 & 27.30 &  1.16& 24.43& 17.90& 12.28&   17& 23.63$\pm$0.04 \\
	&  04 14 35.0& 0.82 & 16.26 & 25.66 & 23.97& 24.02& 16.47& 19.80&   10& 23.04$\pm$0.02 \\
 U8833  &  13 54 48.7& 0.90 & 15.34 & 27.41 &  0.79& 23.71& 15.69& 24.95&   19& 22.42$\pm$0.03 \\
	&  35 50 15.0& 1.50 & 14.64 & 26.50 & 22.81& 23.44& 14.77& 33.66&   11& 21.88$\pm$0.03 \\
 DDO190 &  14 24 43.5& 1.80 & 13.55 & 24.66 &  0.79& 23.32& 13.55& 51.53&   36& 21.78$\pm$0.05 \\
	&  44 31 33.0& 1.85 & 12.77 & 23.96 & 22.26& 22.68& 12.77& 55.58&   22& 21.20$\pm$0.03 \\
	&            &      &       &       &      &      &      &      &     &                \\
SexB    & 10 00 00.1 & 5.10 &       & 24.94 & 0.97 & 23.75& 12.98& 78.41&  64 & 22.60$\pm$0.10 \\
	& 05 19 56.0 & 2.50 &       & 24.02 & 22.69& 22.75& 12.30& 75.24&  38 & 21.50$\pm$0.10 \\
\hline \hline
\end{tabular}
\end{center}
\end{table*}
\end{document}